%% file: main_springer.tex
\RequirePackage{fix-cm}
\documentclass[twocolumn]{svjour3}
\usepackage[T1]{fontenc}
\usepackage[utf8]{inputenc}
\usepackage{mathptmx}
\usepackage{graphicx}
\graphicspath{{./}}
\usepackage{booktabs}
\usepackage{amsmath,amssymb}
\usepackage{multirow}
\usepackage{cite}
\usepackage[hidelinks]{hyperref}
\hypersetup{pdftitle={Candidate Comparability Before Promotion: Conditional Validation in Adaptive Network Intrusion Detection},pdfauthor={Roberto Fern\'andez-Barrios; Iker Pastor-L\'opez; Amaia Pikatza-Huerga; Pablo Garc\'ia Bringas},pdfkeywords={distribution drift; candidate comparability; risk-aware model updating; adaptive model updating; machine learning; intrusion detection}}
\providecommand{\doi}[1]{\url{https://doi.org/#1}}
\usepackage{microtype}
\usepackage[htt]{hyphenat}
\journalname{International Journal of Information Security}
\begin{document}
\title{Candidate Comparability Before Promotion: Conditional Validation in Adaptive Network Intrusion Detection}
\titlerunning{Candidate Comparability Before Promotion}
\author{Roberto Fern\'andez-Barrios \and Iker Pastor-L\'opez \and Amaia Pikatza-Huerga \and Pablo Garc\'ia Bringas}
\authorrunning{R. Fern\'andez-Barrios et al.}
\institute{R. Fern\'andez-Barrios (corresponding author) \and I. Pastor-L\'opez \and A. Pikatza-Huerga \and P. Garc\'ia Bringas \at
  Faculty of Engineering, University of Deusto, Avda. de las Universidades, 24, 48007 Bilbao, Spain\\
  \email{roberto.fernandez.b@deusto.es; iker.pastor@deusto.es; a.pikatza@deusto.es; pablo.garcia.bringas@deusto.es}\\
  ORCID: R. Fern\'andez-Barrios 0009-0003-5312-2634 $\cdot$ I. Pastor-L\'opez 0000-0002-3068-6248 $\cdot$ A. Pikatza-Huerga 0009-0003-9080-6242 $\cdot$ P. Garc\'ia Bringas 0000-0003-3594-9534
}
\date{Received: date / Accepted: date}
\maketitle
\begin{abstract}
Adaptive network intrusion detection systems retrain classifiers after drift alarms, but an alarm detects change; it does not establish that a challenger should replace the deployed incumbent. Promotion is security-relevant because it changes the model responsible for subsequent attack detection, and evaluating it has a methodological problem: promotion conclusions may depend on how the challenger was constructed and on how much evidence supports it. We test that dependence on CICIDS2017, UNSW-NB15 and ToN-IoT with self-contained challenger pipelines, nested candidate-size controls, a common-harness comparison of nine update policies, and a final sensitivity confining every exact feature vector to one evaluation, training or probe role. Incumbent-owned frozen preprocessing amplified apparent promotion harm; with self-contained challenger pipelines the mean full-drift harm did not persist. Raising nominal candidate evidence from 512 to 2{,}000 samples per class improved promotion under pool-constructed progressive drift by $+0.53$, $+1.67$ and $+0.38$ balanced-accuracy points: positive and statistically resolved in all three benchmarks, but materially benchmark-dependent rather than homogeneous, and driven mainly by fewer false positives. Policy conclusions were partially robust: policy ordering changed with candidate comparability, no policy globally dominated, and earlier compatibility statements for a label-free estimator and a calibrated ensemble narrowed. Validation helped evidence-disadvantaged challengers but added no average benefit at parity. Thirteen replays on real, time-ordered traffic showed no net harm from always deploying. Challenger construction and evidence should be controlled, reported and interpreted explicitly when promotion is evaluated.
\end{abstract}
\section{Introduction}

Machine-learning network intrusion detection systems (NIDS) are trained on historical traffic \cite{nguyen2023tsids}, but the traffic they monitor evolves. New services, shifting user behaviour, reconfigured infrastructure and novel attacks move the input distribution away from the training distribution --- concept drift \cite{shyaa2024evolving,wahab2022intrusion,roshan2018adaptive}. A frozen classifier loses accuracy, so operational NIDS periodically \textbf{retrain} on recent traffic \cite{wahab2022intrusion,camarda2025managing,constantinides2019novel}, and the dominant answer to \emph{when} is \textbf{drift detection}: compare recent traffic with a reference window and retrain when a statistical monitor signals change \cite{gama2004learning,rabanser2019failing,lukats2024benchmark}. A large literature therefore competes to build more sensitive monitors, from univariate change detectors to multivariate two-sample tests \cite{gretton2012kernel,szekely2013energy} and quantum-kernel similarity measures \cite{havlicek2019supervised,schuld2019quantum,kaissar2026enhancing}, on the implicit premise that a better change detector yields a better retraining policy.

This paper separates two questions that the detect-then-retrain loop merges. Drift detection answers \emph{has the distribution changed?} Promotion asks \emph{is this particular challenger sufficiently supported to replace this incumbent?} A drift alarm --- or a schedule, or a false alarm --- \emph{proposes} a challenger; it does not establish that the challenger beats the incumbent. In a deployed NIDS that promotion is also a security-relevant integrity decision: it replaces the model responsible for all subsequent attack detection, so a poorly supported update can degrade the detector's future security behaviour even when the drift alarm itself was correct. Between the alarm and the promotion decision sits a layer the loop leaves implicit, and this paper makes it explicit: \emph{candidate comparability}. Before a promotion outcome can be interpreted, let alone automated, the evaluation must establish that incumbent and challenger are comparable in how they were \emph{constructed} and in how much \emph{evidence} supports them. Prior work has framed retraining as a cost--benefit decision rather than an automatic response to detected change \cite{zliobaite2015cost,mahadevan2024cost,regol2025retrain}, and adaptation itself as a choice among mechanisms that includes not adapting \cite{bakirov2021automated}; we add the comparability audit that must precede any such decision, and we show experimentally that it can reverse the conclusions drawn without it.

\textbf{Scope of the net-harm findings.} Several controlled configurations in this paper show always-deploy updating losing to never adapting. These net-harm observations are \emph{configuration-dependent}: they were produced by specific candidate-construction and evidence conditions, and we do not claim them as universal properties of adaptive NIDS or of retraining. The study traces the mean harm to two asymmetries between incumbent and challenger --- a frozen, incumbent-owned preprocessing representation, and a four-fold nominal training-evidence disadvantage --- each isolated by a pre-specified control. Configurations without those asymmetries behave differently: robust learners (random forest, logistic regression, MLP) stay non-negative where the fragile SVC-RBF pipeline goes net-negative, and self-contained, size-matched challengers show no mean deficit at zero drift and a larger benefit under drift. Every harm figure below should therefore be read as a statement about the configuration that produced it, and none of it was observed on real, chronologically ordered traffic: in thirteen chronological replays always-deploy updating never lost to never adapting (\S\ref{sec:boundaries}), which bounds neither the frequency nor the impossibility of such harm in deployment.

The evidence has three layers. An initial controlled study exposed the phenomenon under a historical configuration. Pre-specified replications then isolated each comparability factor with fresh seeds, frozen margins and machine-evaluable outcome rules, extended by a size-matched control under pool-constructed progressive drift and a common-harness comparison with published and reference baselines. A final pre-specified sensitivity repeated the two central blocks with exact-feature-group-disjoint role assignment, after an audit found identical cleaned feature vectors crossing the source-row-disjoint roles of the earlier design. Throughout, \emph{pre-specified} and \emph{preregistered} denote a protocol for an individual stage that was specified and version-controlled before that stage's confirmatory seeds ran; they do not denote an external registry covering the sequential study as a whole.

Building on active-testing and limited-label model-selection research \cite{sawade2012active,kossen2021active,karimi2021online,han2024model,okanovic2025models}, the paper makes three contributions.

\textbf{C1 (primary --- candidate comparability before promotion): promotion conclusions and policy rankings are conditional on challenger construction and evidence.} Adaptive updating is decomposed into \emph{drift trigger} $\to$ \emph{candidate construction} $\to$ \emph{evidence comparability} $\to$ \emph{optional validation} $\to$ \emph{promotion} $\to$ \emph{future deployment outcome} (Fig.~\ref{fig:pipeline}). Each stage answers a distinct question: an alarm proposes; construction and evidence determine the conditions under which the proposal is compared; validation estimates the sign of the specific proposal; and promotion is a decision a healthy incumbent can win. Experimentally, the two upstream conditions changed not only whether promotion appeared harmful or beneficial but also which update policy appeared preferable: in the common-harness comparison the pre-specified ordering-change rule fired for label-free and ensemble alternatives when the challenger's nominal evidence changed, and no evaluated policy dominated globally (\S\ref{sec:baselines}, \S\ref{sec:value_disjoint}). Promotion conclusions should therefore be controlled, reported and interpreted explicitly with respect to challenger construction and evidence.

\textbf{C2 (secondary): controlled identification of two comparability asymmetries.} Preprocessing ownership and nominal candidate-evidence size materially change promotion outcomes. The mean full-drift harm does not persist once each challenger owns its preprocessing (a role-randomized ownership experiment localizes the mechanism to feature scaling; Online Resource~1, \S S2.12); the residual zero-drift mean deficit of 512-per-class challengers disappears at nominal 2{,}000-per-class parity; and under pool-constructed progressive drift raising nominal evidence retains a positive, statistically resolved effect in all three benchmarks under exact-feature-disjoint roles, although the material benefit is benchmark-dependent rather than homogeneous (\S\ref{sec:symmetric}--\S\ref{sec:value_disjoint}).

\textbf{C3 (secondary --- conditional validation and context-dependent policies).} Point/strict validation materially helps when the challenger's construction or evidence is asymmetric and provides no detectable average benefit --- with a measurable strict-gate cost in one cell --- once challengers are self-contained and evidence-matched, at zero drift and under drift alike (\S\ref{sec:gatevalue}). The exact-feature-disjoint policy sensitivity is partially robust: size-dependent ordering and the absence of a global winner survive, while the original ATC and calibrated-ensemble retention statements narrow (\S\ref{sec:baselines}, \S\ref{sec:value_disjoint}).

Supporting instruments and boundaries (tertiary --- formal, diagnostic and operational instruments) complete the study: the sequential VBC-SG family is a secondary formal instrument for probe-level false-superiority control under a stated conditional null; a quantum-kernel monitor serves only as a detector-invariance check; thirteen chronological replays delimit external empirical support --- none showed net harm from always deploying --- without estimating deployment frequency; and an acquisition-yield simulation bounds attack-label discovery cost (\S\ref{sec:boundaries}). These are not universal policies or primary contributions. The paper proceeds: related work (\S2), the decision pipeline and gates (\S3), experimental design (\S4), results (\S5), implications and limits (\S6--\S8).

\section{Related work}

\textbf{Concept drift and adaptive intrusion detection.} Concept drift degrades ML-based NIDS, and a body of work adapts models online or on a schedule to compensate \cite{gama2014survey,lu2019learning,shyaa2024evolving}. Recent systems make that loop more operational. CARAVAN uses labeling agents and an accuracy proxy to trigger online retraining of in-network models, then installs the updated weights; its public artifact covers simulation and an FPGA/P4 testbed \cite{zhang2024caravan}. SPIDER performs semi-supervised continual learning with limited annotations and memory constraints \cite{amalapuram2024spider}, while NOCTOWL incrementally adjusts an interpretable tree under delayed, sampled labels \cite{pederzoli2025noctowl}. SSF fine-tunes continually on a refreshed memory buffer, selecting the new samples that carry the drifted pattern and discarding outdated ones once a Kolmogorov--Smirnov test signals significant drift \cite{zhang2025ssf}. ADAWU-IDS is especially close in motivation: it calibrates response configurations on a held-out chronological validation stream and fixes the selected policy before testing; its selected CICIDS2017 configuration continuously reweights an ensemble and disables hierarchical retraining \cite{yuan2026adawu}. These systems contribute labeling, monitoring, continuous adaptation or stream-level response calibration; none performs the same per-proposal incumbent--challenger promotion comparison, so a direct transplant would change the decision unit and information interface rather than provide a decision-equivalent baseline (Online Resource~1, \S S11). A 2026 journal-level addition to this line is HOIDS, a concept-drift-aware hybrid online intrusion detection system \cite{depaola2026hoids}. We study \emph{whether and when} a concrete retrained candidate should replace its incumbent, connecting to, but distinct from, work on negative transfer and catastrophic forgetting \cite{mccloskey1989catastrophic,kirkpatrick2017overcoming,zhang2023survey}.

\textbf{Drift detection and two-sample monitors.} Drift is commonly detected with change detectors \cite{bifet2007learning,gama2004learning,baena2006early,raab2020reactive}, statistical-test ensembles over batch streams \cite{komorniczak2022ensemble}, or multivariate two-sample tests --- MMD, energy distance, Kolmogorov--Smirnov and Jensen--Shannon divergence --- used as retraining or alarm triggers \cite{gretton2012kernel,rizzo2016energy,lin1991divergence,darling1957ks}; how drift manifests (type, locality, severity) itself varies widely across streams \cite{aguiar2024locality}. Our classical baselines are drawn from this family. A complementary family of \emph{performance-aware} detectors monitors the model's error signal directly, presupposing continuous labeled feedback; Bayram et al. \cite{bayram2022degradation} formalize the distinction between distributional change and model degradation that our degradation--headroom analysis (Online Resource~1, \S S1.1) quantifies empirically. Improving distribution-level monitors did not improve the promotion decision in any regime we evaluated, because they are blind to model degradation; the reference DDM and ADWIN implementations are evaluated as retraining triggers in the common-harness comparison (\S\ref{sec:baselines}).

\textbf{Quantum kernels for drift monitoring.} Quantum-kernel methods embed data via parameterized quantum feature maps \cite{havlicek2019supervised,schuld2019quantum,huang2021power}, with exploratory use in cybersecurity \cite{gouveia2020towards,nicesio2023quantum,bellante2025evaluating}; a quantum-kernel MMD monitor is included here only as a detector-invariance check. Changing the detector family did not change any promotion outcome we evaluated, consistent with evidence that quantum-kernel advantage claims require careful classical baselining and face concentration limits \cite{schnabel2025quantum,thanasilp2024concentration}.

\textbf{Drift, rejection and evaluation discipline in ML security.} A parallel security literature asks when a deployed classifier should no longer be trusted. Transcend \cite{jordaney2017transcend} and its revision Transcendent \cite{barbero2022transcendent} use conformal evaluation to reject individual predictions whose low credibility signals model aging; CADE \cite{yang2021cade} detects and explains drifting samples via contrastive representations; INSOMNIA \cite{andresini2021insomnia} updates a NIDS continuously with semi-supervision as drift accrues. These systems abstain on individual samples or adapt continuously; none evaluates whether a specific retrained candidate should replace the incumbent, and none can be instantiated in this decision problem without changing what it decides (\S\ref{sec:baselines}). Sommer and Paxson \cite{sommer2010outside} documented early the striking gap between extensive academic research on ML-based anomaly detection and its rare operational deployment, naming fundamental difficulties of sound evaluation among the causes; later work sharpened that concern through temporal, dataset and security-specific evaluation pitfalls: TESSERACT \cite{pendlebury2019tesseract} and the ``dos and don'ts'' of Arp et al. \cite{arp2022dodonts} document how temporal and experimental bias inflate security-ML results; Apruzzese et al. \cite{apruzzese2023sok} systematize the gap between benchmark performance and a pragmatic assessment of ML-based NIDS, and Jacobs et al. \cite{jacobs2022emperor} show that benchmark-accurate network-security models can rest on shortcuts that do not survive a change of conditions. Closest to the update decision itself, Kan et al. \cite{kan2021labelless} show a self-updating malware detector degrading as unverified pseudo-labels feed its own retraining, and Chen et al. \cite{chen2023continuous} show that continuous retraining under drift depends on which samples are labeled and how; both concern the update rather than the alarm. Our pre-specified protocol with temporally ordered progressive windows and paired within-run baselines (\S4) follows that evaluation discipline.

\textbf{When to retrain: cost-aware and validation-gated updates.} Beyond detection, some work weighs the cost of adaptation or validates a candidate before deployment, e.g. in MLOps model-update pipelines and shadow/champion--challenger evaluation \cite{nath2007champion,sculley2015hidden,kreuzberger2023mlops,guissouma2023continuous}; that a well-intentioned update can be net-harmful is documented beyond security \cite{bansal2019updates}. More generally, automated stream-learning frameworks separate prediction, made with the current model, from the subsequent adaptation step and treat adaptation as a choice among multiple mechanisms, including no update; Bakirov et al. \cite{bakirov2021automated} select the mechanism by cross-validation on the most recent batch. Our decision unit is narrower: conditional on a NIDS update proposal, we compare a concrete challenger against the deployed incumbent while controlling how that challenger was constructed and what evidence it received. Champion--challenger evaluation, validation gates and label-efficient model comparison are therefore not new, and this paper claims none of them as novel; their reliance on a small labeled probe connects to active and label-efficient learning for data streams \cite{cacciarelli2024active,krawczyk2019adaptive,liu2021comprehensive}. A complementary line estimates a model's accuracy under distribution shift without target labels \cite{garg2022atc,guillory2021doc,deng2021labels,baek2022agreement,jiang2022disagreement,rosenfeld2023dis2}; we instantiate its two standard estimators (ATC, DoC) as promotion rules under the same harness as our gates (\S\ref{sec:baselines}).

\textbf{Positioning and novelty.} Table~\ref{tab:positioning} positions the closest decision formulations by objective, decision unit, required information, output and guarantee. The rows are complementary rather than an empirical ranking: their problem definitions, information sets and outputs differ, and this paper reports no head-to-head superiority claim against them. What this study adds is not a new gate but an experimentally supported \emph{decomposition} of adaptive promotion: adaptive-NIDS systems increasingly expose labeling, monitoring, updating and calibration interfaces \cite{zhang2024caravan,amalapuram2024spider,pederzoli2025noctowl,yuan2026adawu}, whereas we separate promotion from the alarm and show, with pre-specified controls, that upstream challenger construction and evidence conditions change both the apparent value of promotion and the apparent relative value of update policies. Comparability of candidates is a familiar concern in model comparison generally \cite{sawade2012active,karimi2021online}; we claim only that its consequences for this per-proposal decision had not been isolated under the controls used here. Prior work distinguishes drift monitoring from adaptation or retraining decisions \cite{bayram2022degradation,regol2025retrain,bakirov2021automated} and has used validation to calibrate adaptive responses \cite{yuan2026adawu}, but, to our knowledge, it does not isolate challenger comparability as a factor governing per-proposal incumbent--challenger promotion in adaptive NIDS, and the value of validation as a per-proposal promotion gate under controlled challenger-comparability regimes remains underexplored.

\begin{table*}[t]
\centering
\caption{Conceptual positioning of related decision formulations. ``Guarantee'' records the type and scope stated by each formulation, not a common empirical ranking.}
\label{tab:positioning}
\scriptsize
\setlength{\tabcolsep}{2pt}
\renewcommand{\arraystretch}{1.12}
\begin{tabular}{p{0.12\linewidth} p{0.16\linewidth} p{0.12\linewidth} p{0.18\linewidth} p{0.13\linewidth} p{0.20\linewidth}}
\toprule
Approach & Objective & Decision unit & Information required & Output & Guarantees / scope \\
\midrule
CARA \cite{mahadevan2024cost} & Cost-aware stream retraining & Retraining opportunity & Performance and retraining/staleness costs & Retrain or keep & Cost--performance analysis under its assumptions; no concrete-challenger audit \\
Regol \cite{regol2025retrain} & Forecast when retraining pays & Retraining opportunity & Sparse performance history and cost ratio & Retrain or keep & Uncertainty-aware forecast within its model; no paired proposal guarantee \\
Automated adaptation strategies \cite{bakirov2021automated} & Select among stream adaptation mechanisms & Batch / adaptation opportunity & Recent-batch (cross-validation) evidence & Selected adaptive mechanism, including no update & Adaptive-strategy selection within its framework; no explicit incumbent--challenger comparability audit \\
IGPC-MSOS \cite{shyaa2026igpcmsos} & Select NIDS update mode & Drift/update episode & Dual triggers and online system feedback & Update mode & System-specific decision architecture; no paired concrete-challenger guarantee \\
CARAVAN \cite{zhang2024caravan} & Label, monitor and retrain in-network models & Labeling window & Generated labels, predictions and accuracy proxy & Retrain/update weights & End-to-end adaptive system and public artifact; no held-out challenger comparison \\
ADAWU-IDS \cite{yuan2026adawu} & Calibrate drift-response policy & Validation stream / traffic chunk & Delayed chunk labels, drift severity and learner performance & Reweight; optional stronger response & Chronological held-out calibration; selected configuration is dataset-specific \\
Active comparison/testing \cite{sawade2012active,kossen2021active} & Compare fixed predictors label-efficiently & Query/evaluation round & Unlabeled pool and queried labels & Estimate or ranking & Method-specific sample-efficiency/statistical guarantees \\
Limited-label model selection \cite{karimi2021online,han2024model,okanovic2025models} & Select among available models & Selection round/domain & Candidate predictions, labels and/or history & Selected model & Regret or selection guarantees under stated assumptions \\
This work & Audit comparability; condition promotion & Drift-triggered proposal & Incumbent/challenger pipelines, training evidence and a small probe & Commit, reject or defer & Comparability evidence is central; VBC-SG secondarily controls false probe-superiority, not future accuracy \\
\bottomrule
\end{tabular}
\end{table*}

\section{Candidate promotion as a decision pipeline}

\subsection{Promotion as a sequential decision}

We consider a deployed binary classifier $h_0$ (benign vs. attack) monitoring a stream of traffic windows $W_1, W_2, \dots$, each a set of flows with (at evaluation time) ground-truth labels; the deployed model degrades as the traffic distribution drifts. A drift monitor $D$ produces a score $s_t = D(W_t)$ measuring divergence of $W_t$ from a reference, and the standard adaptive-NIDS loop raises an alarm when $s_t$ exceeds a threshold and \textbf{retrains} on recent data upon a confirmed alarm. This loop conflates two questions:

\begin{itemize}
  \item \textbf{detection:} \emph{has the distribution changed?} --- answered by $D$;
  \item \textbf{decision:} \emph{will replacing $h$ with a model retrained on the current window improve accuracy?} --- which $D$ does not answer.
\end{itemize}

Let $h$ be the incumbent model at proposal time $t$ and $h'$ the \emph{challenger}: a model retrained on the current regime and proposed for promotion (we use \emph{challenger} for the model itself and \emph{candidate} for its role in the promotion decision, as in \emph{candidate construction} and \emph{candidate evidence}). For a deployment horizon $H$, define the candidate's realized future value relative to retaining the incumbent as
\[
\begin{aligned}
V_t(h',h;H)
  &=\sum_{k=1}^{H} w_k\!\left[\mathrm{BA}(h',W_{t+k})-\mathrm{BA}(h,W_{t+k})\right],\\
w_k&\geq0,\qquad \sum_{k=1}^{H}w_k=1 .
\end{aligned}
\]
Promotion is beneficial over that horizon iff $V_t(h',h;H)>0$ (or exceeds a deployment-cost equivalent expressed in the same units). This is a future, generally unobserved quantity at decision time. The labeled probe instead supplies a contemporaneous estimate of the sign of the candidate--incumbent difference and is evaluated here as a surrogate promotion rule; it is not an observation of $V_t$.

\begin{figure*}
\centering
\includegraphics[width=\textwidth]{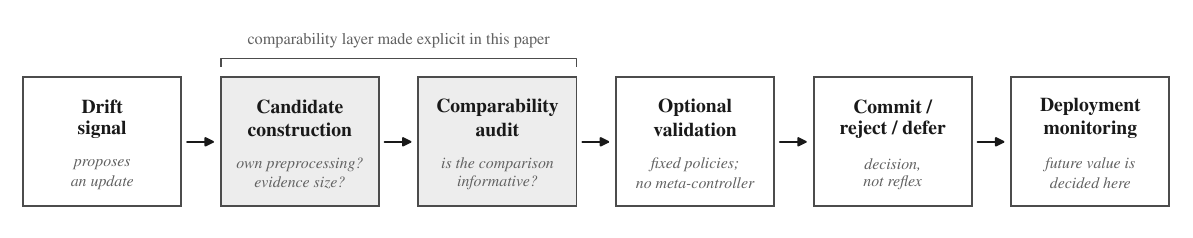}
\caption{Candidate promotion as a decision pipeline. A drift alarm proposes, but does not justify, promotion; candidate construction and evidence comparability precede candidate validation, and the validation stage is optional --- conditional on construction, evidence and incumbent health --- rather than universal.}
\label{fig:pipeline}
\end{figure*}

Because $D$ measures distributional change rather than the sign of this relative value, a detector-only policy commits every confirmed change, including changes for which retraining does not help. The pipeline of Fig.~\ref{fig:pipeline} separates the alarm, the proposal, the evidence used to \emph{train} the candidate, the evidence used to \emph{validate} it, and the evidence that determines its \emph{future deployment} value --- five objects the detection-centric loop treats as one. The gates of \S\ref{sec:gates} instantiate the validation stage; the controls of \S\ref{sec:construction} instantiate the construction and comparability stages.

\subsection{Drift monitors}

We use two-sample drift monitors comparing a reference window (from the pre-drift regime) to the current window: \textbf{Energy distance}, \textbf{RBF-MMD}, per-feature \textbf{Kolmogorov--Smirnov} (max reduction) and histogram \textbf{Jensen--Shannon} divergence, plus classically simulated \textbf{quantum-kernel MMD} monitors with \textbf{ZZ} and \textbf{Pauli-XZ} feature maps used only for the detector-invariance check of \S\ref{sec:v2} (Online Resource~1, \S S1.2). All monitors expose a scalar score; thresholds are calibrated per detector on held-out calibration windows at a fixed quantile (0.95). The paper's claims are detector-agnostic.

\subsection{Pool-constructed progressive-drift readaptation protocol}

Each stream has 100 post-drift windows. Window $t$ is a class-balanced sample in which each class is a mixture of empirical reference and current regime pools with mixing weight $\mathrm{sev}(t)$ ("severity"), ramped linearly from 0 to 1 over the first 80 windows and held at 1 thereafter. We call this \emph{pool-constructed progressive drift}: the source regimes are empirical, but the temporal trajectory is constructed. On a committed adaptation, the candidate is trained on a fresh balanced labeled sample from the current regime, and the detector reference is reset to the adapted regime. The trigger policy (alarm confirmation and cooldown) is held fixed across all gate conditions, so the decision gate is the only \emph{designed} difference between arms. On the confirmatory harness every arm additionally processes \emph{bit-identical} streams, so contrasts are exactly paired; only the initial exploratory study drew a separate pseudo-random realization per arm (identical in distribution, not in realization; \S4).

\subsection{The validate-before-commit gate}
\label{sec:gates}

The gate is a decision step interposed between the trigger and the model swap: a \emph{deployment} (commit) gate, not a retraining gate. On each confirmed trigger the loop retrains a candidate $h'$ as usual --- consuming its training labels and compute regardless of the outcome --- and the gate then decides only whether $h'$ replaces the incumbent $h$; otherwise we keep $h$ (and keep the detector reference, so a later window can re-propose once an update becomes worthwhile). \S\ref{sec:v2} additionally evaluates a two-stage variant that moves the same probe \emph{before} candidate training, gating the training decision itself. We evaluate three commit gates:

\begin{itemize}
  \item \textbf{always-deploy} (\texttt{none}; written \emph{naive} in estimand names and tables): always commit --- the standard loop.
  \item \textbf{point gate} (\texttt{labeled\_probe}$(b,\varepsilon)$): draw a small balanced labeled probe $P$ of $b$ flows from the current window; commit iff $\mathrm{BA}(h',P) \ge \mathrm{BA}(h,P) + \varepsilon$. The probe is the method's incremental labeling cost; we report $b\in\{32,64\}$, $\varepsilon=0$. The \textbf{strict gate} is the same rule rejecting probe ties.
  \item \textbf{disagreement gate} (\texttt{unsup\_disagree}$(\tau)$): commit iff the fraction of flows on which $h$ and $h'$ disagree on the \emph{unlabeled} current window is $\ge \tau$ (a zero-label proxy for "adaptation would change decisions"); we report $\tau=0.15$.
\end{itemize}

Two harness variants, which differ in exactly the information they may use, are stated in Fig.~\ref{fig:algorithm}. Variant~A is the \emph{core simulation} gate: the environment supplies candidate batches and probes from its class pools at the current mixture severity, the source of the main tables but an oracle a deployment lacks. Variant~B is the \emph{observed-data} gate: the candidate's training rows, the probe and the detector's post-commit reference all come from windows the system has actually seen, and nothing reads $\mathrm{sev}(t)$; \S\ref{sec:v2} reports B as a replication of A.

\begin{figure*}[t]
\small
\begin{verbatim}
(A) Gated update, core simulation harness (one stream)
  h <- h0 ; D <- fit_reference(pools, sev=0) ; alarms <- [] ; cooldown <- 0
  for t = 1..T:
      W_t <- environment.stream[t]              # pre-generated, balanced, labeled
      record BA(h, W_t)
      s_t <- D.score(W_t) ; alarms.append(s_t > threshold)
      trigger <- policy(alarms) and cooldown == 0
      if trigger:
          h' <- train(pools.train, sev(t))      # candidate: current mixture (oracle)
          P   <- sample(pools.probe, sev(t))    # probe: balanced, b flows (oracle)
          commit <- [ BA(h',P) >= BA(h,P) + eps ]                  # Sec 3.4
          if commit:
              h <- h' ; D <- fit_reference(pools, sev(t))          # deploy + reset
          alarms <- [] ; cooldown <- C                             # applied either way
      else: cooldown <- max(0, cooldown - 1)
  return {BA(h, W_t)}_t

(B) Gated update, observed-data harness (no sev(t), no pools)
  ... identical control flow, but at a trigger at window t:
      h' <- train(rows of the last 8 OBSERVED windows)             # candidate
      P  <- 32 rows of the OBSERVED window t-9, natural composition,
            excluding by row identity any row used to train h'     # probe
      commit <- [ acc(h',P) >= acc(h,P) + eps ]
      if commit: h <- h' ; D <- fit_reference(last 8 OBSERVED windows),
                 threshold <- quantile_0.95(scores of preceding OBSERVED windows)
\end{verbatim}
\caption{The gated update loop. (A) Core simulation harness: candidate batches and probes are drawn from the environment's class pools at the current mixture severity (an oracle a deployment lacks; source of the main tables). (B) Observed-data harness: candidate, probe and detector recalibration come only from windows the system has already seen, and nothing reads the severity. The commit rule is that of \S\ref{sec:gates}.}
\label{fig:algorithm}
\end{figure*}

The point gate estimates the contemporaneous sign of the candidate--incumbent difference from $b$ labels, the surrogate for future value identified in \S3.1; the disagreement gate tests whether that estimate can be obtained without labels.

\textbf{A cost-sensitive reading: two decisions, two thresholds.} Let each error cost $\lambda_e$ per flow, deploying an update $\lambda_U$, and training a candidate $\lambda_C$ (compute plus training labels). \emph{At commit time the candidate already exists}: $\lambda_C$ and the probe are sunk, so over a horizon of $N$ post-decision flows the myopic commit rule is
$$\mathrm{commit} \iff \widehat{\mathrm{BA}}(h',P) - \widehat{\mathrm{BA}}(h,P) \ge \varepsilon^{\ast}, \qquad \varepsilon^{\ast} = \frac{\lambda_U}{\lambda_e N},$$
--- the margin $\varepsilon$ is the \emph{deployment} cost in accuracy points over the remaining horizon; $\lambda_C$ does not belong in it, because rejecting the candidate cannot recover it. Training costs enter one decision earlier, whether to build a candidate at all, which the two-stage gate of \S\ref{sec:v2} instantiates by spending the probe on the incumbent \emph{before} training. A risk-averse operator replaces the point estimate with a lower confidence bound (the LCB variant of \S\ref{sec:v2}); raising $\varepsilon$ from 0 to 0.01--0.02 cuts committed updates by ${\sim}30\%$ at equal downstream gain (Online Resource~1, \S S1.5), and asymmetric error costs substitute a weighted accuracy for BA in the same rule.

\subsection{Risk-controlled commit gates and the scope of their guarantee}
\label{sec:riskgates}

This subsection states the risk-controlled alternatives to the point gate at the level needed to read the results; implementation details, the deferral continuation modes, the spending schedules and the stated properties are in Online Resource~1, \S S2.13, and the proof of Proposition~1 in \S S4. VBC-SG is a secondary decision-support instrument for settings in which validation is chosen; it is not the source of the paper's main conclusion, and we claim neither that any member of the family dominates nor mathematical novelty for the construction.

\textbf{(A) Point and strict validation.} The point gate ($\varepsilon{=}0$) commits on any observed probe advantage; it maximizes the probe point estimate, not future accuracy, and carries no formal error control: in the logged-trigger analysis of the historical configuration, 23\% of its commits had negative realized value (Online Resource~1, \S S2.10). The \emph{strict} variant additionally rejects probe ties at no additional labels, the better buy at some operating points (\S\ref{sec:risk}).

\textbf{(B) Risk-controlled validation.} Let the probe be $b$ flows drawn from a distribution $\mathcal P$, and for each flow let
$$d_i = \mathbf 1[h'\text{ correct on }i] - \mathbf 1[h\text{ correct on }i]\in\{-1,0,+1\}.$$
Three rules commit only on statistical evidence that $\mathbb E_{\mathcal P}[d]>0$: (i) \textbf{exact McNemar} \cite{mcnemar1947note}, one-sided on the discordant pairs at level $\alpha$ (ties never commit); (ii) a \textbf{Robbins} normal-mixture lower confidence sequence (CS) \cite{howard2021timeuniform}; (iii) an \textbf{empirical-Bernstein} betting confidence sequence \cite{waudbysmith2023betting}, much tighter because the $d_i$ are mostly ties. The CS gates inspect the probe in 16-flow blocks up to the budget $b$ and reject any proposal whose lower bound never clears $0$; they run at $\alpha=0.10$, exact McNemar at its pre-specified $\alpha=0.05$ (McNemar at $0.10$ leaves the zero-drift conclusion unchanged; \S\ref{sec:risk}).

\textbf{(C) Pooled versus stratified guarantees.} The probe is drawn with \emph{fixed class quotas} ($b/2$ per class), not i.i.d.\ from one distribution, so the pooled tests above match the sampling only approximately. \emph{Stratified} variants lower-bound each class's mean correctness difference at $\alpha/2$ (Bonferroni over the two classes) and commit iff $\tfrac12(L_{\mathrm{ben}}+L_{\mathrm{att}})>0$, which targets the \emph{balanced-accuracy} difference the design estimates; each per-class bound inherits the guarantee of (E) under that class's conditional null. Only the anytime-valid stratified version, in which each class's bound is an empirical-Bernstein confidence sequence, supports the anytime-valid claim; a fixed-sample per-class LCB variant has no sequential validity.

\textbf{(D) VBC-SG: commit, reject or defer.} The \emph{Validate-Before-Commit Sequential Gate} carries every layer of control at once. Per-class empirical-Bernstein sequences drive a three-action decision: \textsc{commit} when the balanced-accuracy lower bound clears zero, \textsc{reject} when the upper bound falls below zero (futility), and \textsc{defer} otherwise, retaining the candidate and continuing at later windows up to a deferral cap $D$. Because the stream can drift between windows, a deferred sequence must state what it estimates; three continuation modes do so (\emph{Cohort-sim}, \emph{Refresh} and \emph{Accumulate}; Online Resource~1, \S S2.13). Cohort-sim resamples the proposal-time target in the simulator and is not a retained production cohort, so that acquisition scheme is not claimed to be directly deployable. A deployment-long risk budget allocates each proposal's level $\alpha_j$ by a declared spending schedule (Bonferroni or $p$-series).

\textbf{(E) What VBC-SG guarantees.} For a single proposal and \emph{under the probe distribution} $\mathcal P$, the confidence-sequence gates satisfy
$$\Pr\!\big(\exists n\le b:\ \text{LCB}_n>0 \ \big|\ \mathbb E_{\mathcal P}[d]\le 0\big)\ \le\ \alpha ,$$
uniformly over the data-dependent stopping time, \emph{when the conditional weak null of Proposition~1 holds along the adjudication sequence}: they control the probability of \emph{falsely establishing positive expected paired correctness on the probe distribution}. Under the spending schedule the union bound gives a deployment-long version (a false commit \emph{ever} occurs with probability at most $\alpha_{\mathrm{life}}$); without a schedule the guarantee is per proposal only, and labels per proposal are bounded by the per-window budget times $(1+D)$. Three things should be kept apart. The \emph{theorem} is the anytime-valid statement below under its conditional weak null. The \emph{implementation} applies the confidence sequence separately by class in the stratified gates. The \emph{limitation} is that the probe is drawn with fixed finite class quotas (and, in the pool harness, a fixed reference/current mixture) before random ordering, and it is not proved here that a marginal non-superiority null $\mathbb E_{\mathcal P}[d]\le 0$ under such quotas implies the sequential conditional null the theorem needs; the applied guarantee is therefore stated only under that conditional null.

\textbf{Proposition 1 (conditional validity of the commit e-process).} Map each per-flow correctness difference to $x_i = (d_i+1)/2 \in [0,1]$; the empirical-Bernstein sequence implemented in the artifact is a wealth process $W_n$ with predictable betting fractions and plug-in means, whose commit condition $\mathrm{LCB}_n > 0$ is algebraically the event $W_n > 1/\alpha$. If $\mathbb{E}[d_i \mid \mathcal F_{i-1}] \le 0$ for every $i$ --- the weak null; no stationarity or identical distribution is assumed --- then $W_n$ is a nonnegative supermartingale with $W_0 = 1$, and Ville's inequality gives $\Pr(\exists n : W_n > 1/\alpha) \le \alpha$. The proof (Online Resource~1, \S S4) is the standard testing-by-betting argument \cite{waudbysmith2023betting,howard2021timeuniform}; we claim no novelty, only the correspondence to this implementation, which a null simulation with time-varying conditional means and variances exercises empirically (a check of the implementation under the conditional null, not a proof for the quota sampling design).

\textbf{(F) What VBC-SG does not guarantee.} None of the above bounds the probability that a committed candidate lowers \emph{future} balanced accuracy: the probe-level bound and a future-deployment bound coincide only when the probe is representative of the windows the candidate will face, and the stale-probe and dependence results of \S\ref{sec:boundaries} show where that premise fails. We therefore claim a bound on \emph{false probe-superiority} under the stated conditional null, never on future deployment regret or harm, and never a defense against a manipulated probe; \S\ref{sec:risk} reports the empirical harmful-commit accounting and the label price of each layer.

\section{Experimental design}

\subsection{Datasets and controlled streams}

\textbf{Datasets and regimes.} CICIDS2017 \cite{sharafaldin2018toward} (Tuesday reference $\to$ Wednesday, Friday-DDoS, Friday-PortScan, Thursday-WebAttacks), UNSW-NB15 \cite{moustafa2015unsw} (DoS, Reconnaissance) and ToN-IoT \cite{alsaedi2020toniot} (Scanning), spanning benefit, marginal/mixed and harmful readaptation. Features are numeric, standardized and PCA-reduced to 8 dimensions; windows contain 128 flows. To rule out label leakage, identifier and label-derived columns are removed at staging (UNSW: \texttt{id}, \texttt{label}, \texttt{attack\_cat}; ToN-IoT: \texttt{label}, \texttt{type}, timestamps and IP/uid identifiers; the CICIDS variant used ships no flow identifiers), and no remaining single feature separates the classes beyond AUC $\approx$ 0.81. CICIDS2017 has documented traffic-generation and labeling artifacts \cite{engelen2021troubleshooting,lanvin2022errors}, and benchmark NIDS datasets more generally carry design smells such as heavy duplication and unclear labeling \cite{flood2024smells}; the concern is not specific to CICIDS2017, since class imbalance and data overlap are documented for UNSW-NB15 \cite{zoghi2024unsw}, and a recent cross-dataset review compiles published duplication, cross-split contamination, overlap and label-inconsistency figures for CICIDS2017, UNSW-NB15 and ToN-IoT, obtained under heterogeneous protocols \cite{aburomman2026automl}. The exact-feature-disjoint sensitivity of \S\ref{sec:value_disjoint_method} addresses exact duplicate-value exposure across evaluation, training and probe roles directly, under its own reproducible definition rather than these sources' overlap measures; we therefore treat CICIDS2017 as the benefit end of a three-benchmark spectrum rather than as sole evidence --- the paper's central negative results (the marginal and harmful regimes) come from the independent UNSW-NB15 and ToN-IoT benchmarks. The exploratory stage spans all seven regimes; every registered confirmatory block (\S\ref{sec:construction}--\S\ref{sec:baselines_method}, and the historical replication of \S\ref{sec:historical}) uses exactly one primary scenario per benchmark --- CICIDS2017 PortScan, UNSW-NB15 Reconnaissance and ToN-IoT Scanning --- under full drift and, where stated, zero drift.

\textbf{The confirmatory harness.} All confirmatory results come from a harness built to make the paired contrasts exact. (i) \emph{Common streams:} every evaluation stream and calibration draw is pre-generated from the seed alone before any policy runs, so all arms process bit-identical windows and policy actions cannot perturb the environment. (ii) \emph{Source-row-disjoint partitions:} each per-class pool is split once per seed (50/30/20) into window/train/probe partitions used respectively for evaluation windows, for initial models, candidates and detector references, and for probes; a source-row index cannot cross roles, but distinct source rows carrying an identical cleaned feature vector can, so these blocks are not exact-feature-value-disjoint. (iii) \emph{Separate generators:} environment, initial training, detector calibration, candidate training, probe draws and post-commit recalibration each use their own RNG, seeded by (seed) or (seed, $t$) only. (iv) \emph{Per-trigger logging:} at every trigger the harness records pre-trigger incumbent degradation, the detector score, and incumbent-vs-candidate BA over the following 1/3/5/10 windows, lookahead used for logging only. The exact-feature-disjoint sensitivity changes only item (ii).

The replication protocol (criteria, margins, one-pass stopping rule) was committed and publicly tagged in version control before any confirmatory seed ran; its confirmatory window is 30 held-out seeds (104--133), untouched by any smoke test.

\textbf{Shared protocol constants.} The initial model trains on 2{,}000 flows per class from the reference regime; in the historical configuration the scaler and PCA are fit on that training set only and frozen thereafter, and every candidate trains on a balanced sample of 512 flows per class drawn \emph{at the current mixture severity} $\mathrm{sev}(t)$, never from the final regime pool: the candidate sees the population the deployed model faces at the trigger, a simulator convenience a production system lacks (\S7; the observed-data and chronological arms are the counterpoints). The comparability controls of \S\ref{sec:construction} vary exactly these two conventions. Detector references use 256 flows per class; thresholds are the 0.95 quantile of detector scores over 30 pre-drift calibration windows; the trigger is 3 consecutive alarms with a 10-window cooldown; windows contain 128 flows. The probe is the gate's \emph{incremental} label cost on top of the labeled retraining data any labeled-retraining loop already consumes: we claim label efficiency of the \emph{commit decision}, not of the retraining pipeline (\S\ref{sec:v2} accounts for both).

\textbf{Label and information latency.} Several arms model the delay between a flow being observed and its label being available; Table~\ref{tab:latency} fixes the semantics, read directly off the released code. The probe knob ($L_p$) draws the probe from the stream as of window $t-L_p$; the candidate knob ($L_c$) shifts the candidate's \emph{whole} training batch, features and labels together, to the mixture of window $t-L_c$ (pool generator) or the eight observed windows ending at $t-L_c$ (sliding-window generator). The observed-data probe is inherently nine windows old. Training completion is instantaneous except in the operational sensitivity (\S\ref{sec:v2}), which evaluates a five-window deployment delay while the incumbent keeps serving. While a deferred VBC-SG proposal is pending, the incumbent remains deployed and \emph{new proposals are blocked}, exactly like a cooldown window, so a proposal is never silently replaced.

\begin{table*}[t]
\centering
\caption{Temporal (causal-information) availability of every object the update loop touches at a trigger at window $t$. ``Latency knob'' names the runner flag; 0 = contemporaneous.}
\label{tab:latency}
\footnotesize
\setlength{\tabcolsep}{3.5pt}
\begin{tabular}{l l l l}
\toprule
Object & Source & Reflects window & Latency knob \\
\midrule
Trigger decision & detector score on $W_t$ & $t$ & --- \\
Probe (pool arms) & pool sample at $\mathrm{sev}(t-L_p)$ & $t-L_p$ & \texttt{--probe-latency} \\
Probe (observed arm) & observed window $t-9$, row-disjoint & $t-9$ & fixed \\
Candidate batch (pool) & pool sample at $\mathrm{sev}(t-L_c)$ & $t-L_c$ & \texttt{--candidate-latency} \\
Candidate batch (sliding) & 8 observed windows ending $t-L_c$ & $[t-L_c-7,\, t-L_c]$ & \texttt{--candidate-latency} \\
Recalibration (observed) & last 8 observed windows & $\le t$ & --- \\
Training completion & instant; delayed only in the operational arm & $t$ / $t{+}5$ & \texttt{--training-delay} $\in\{0,5\}$ \\
Deferred proposal & pending state; blocks new proposals & until resolved & \texttt{--defer-windows} \\
\bottomrule
\end{tabular}
\end{table*}

\textbf{The initial exploratory study.} The hypotheses, regime taxonomy and robustness texture of Online Resource~1, \S S1 come from an earlier harness whose arms drew seed-matched but unpaired stream realizations. Its numbers are exploratory context: every claim the paper rests on was re-established on the hardened harness with fresh seeds, and exploratory rankings never govern where confirmatory evidence exists. A pre-specified confirmation/cooldown policy grid with no decision gate is a negative baseline (Online Resource~1, \S S3).

\subsection{Candidate construction and evidence controls}
\label{sec:construction}

We evaluate how preprocessing ownership and candidate evidence size alter promotion outcomes through staged, pre-specified controls: each factor was isolated by a replication whose protocol, estimands, margins and interpretation rules were frozen before any of its confirmatory seeds ran. The controls were designed sequentially, each specified after the previous one's outcome was known; we present them as the two levels of a construction factor and two drift regimes of an evidence factor rather than as a chronology.

\subsubsection{Transformer ownership: self-contained candidate pipelines}
\label{sec:symmetric_method}

The historical configuration of this study's harness fits scaler and PCA once, on the initial training set, and \emph{freezes} them --- every later challenger is trained and scored inside that initial representation. A role-randomized ownership control (Online Resource~1, \S S2.12) shows this policy is not neutral for SVC-RBF --- giving the incumbent only the feature standardizer reproduces essentially the whole ownership advantage --- so the decisive question is dynamic: does the harmful-update result survive when each challenger is a \emph{self-contained pipeline}? A pre-specified replication answers it (protocol archived with the public artifact). Throughout, ``historical'' denotes this study's own earlier harness: we do not claim that frozen incumbent-owned preprocessing is a standard or widespread policy in published adaptive NIDS; the lesson concerns asymmetric construction in \emph{evaluation}, not the prevalence of the policy.

\textbf{Design.} The environment emits a \emph{raw} stream, hash-verified bit-identical across arms per seed. Each model is a self-contained pipeline (scaler + PCA + classifier) that transforms internally; the drift monitor keeps the initial transformer under \emph{both} policies, so construction policy is never confounded with monitoring policy. The policies differ in exactly one intervention: \texttt{frozen\_initial\_transformer} reuses the initial scaler/PCA for every candidate (reproducing the historical harness bit-for-bit), while \texttt{own\_transformer\_per\_model} fits each challenger's scaler and PCA on its own candidate batch only, with the same dimension, solver and classifier hyperparameters; raw candidate batches are bit-identical across policies, the probe is served raw to both, and neither probe nor future windows enter any fit. A commit deploys the complete bundle from window $t{+}1$; scaler/PCA hashes and the SVC's fitted $\gamma$ are recorded per candidate.

\textbf{Estimands and inference.} Fresh confirmatory seeds 3001--3030, six scenarios (PortScan, UNSW-Recon, ToN-IoT $\times$ full and zero drift), KS-max only, policies never-adapt/always-deploy/point/strict. The pre-specified estimands are $\Delta_{\text{naive,own}} = \mathrm{BA}_{\text{naive,own}} - \mathrm{BA}_{\text{never}}$ (does harm persist?), $\Delta_{\text{transformer}} = \mathrm{BA}_{\text{naive,own}} - \mathrm{BA}_{\text{naive,frozen}}$ (the ownership interaction, which isolates the ownership effect directly), and $\Delta_{\text{point--naive,own}}$, $\Delta_{\text{strict--naive,own}}$ (does validation retain value?). Four frozen families (harm under own/full; transformer interaction/full; gate value under own/full; a 12-contrast zero-drift secondary family) are tested by the deterministic centered paired bootstrap used throughout, Holm-corrected within family; the inferential unit is the \emph{seed}, and windows, triggers and commits are never treated as independent units.

Margins were frozen in advance: $\pm0.5$ BA points for equivalence, assessed by inclusion of the CI90 within the margin (the interval-inclusion form of the two one-sided tests \cite{schuirmann1987comparison}; $\pm0.2/\pm1.0$ sensitivities), and non-inferiority guardrails of $-1.0$ recall points and $+0.5$ FPR points (one-sided 95\% bounds) that restrict safety \emph{language}; balanced accuracy alone determines the pre-specified verdict (Scenario A, B or C). The $\pm0.5$-point margin equals the materiality threshold this study pre-specified for its own harm claims (effects below half a point are treated as operationally immaterial; the smallest harm called material, $-0.65$, exceeds it), so CI90 inclusion records compatibility of the mean effect with that margin. It is a study-level preregistered materiality and equivalence threshold, not a universal operational security threshold; every equivalence conclusion remains margin-dependent and does not establish absence of an effect.

\subsubsection{Candidate evidence size at zero drift: the size-matched control}
\label{sec:sizematched_method}

The replication above trains every challenger on 512 flows per class against a 2{,}000-per-class incumbent, so its residual zero-drift harm admits one mundane explanation: the challenger sees a quarter of the incumbent's evidence. A pre-specified control (protocol, config, seeds, margins, families and machine-evaluable outcome rules committed before any confirmatory seed ran) isolates that variable: does harmful promotion under zero drift persist when self-contained challengers are trained with the \emph{same} per-class sample size as the incumbent?

\textbf{Design.} Zero drift only, with the symmetric replication's scenario definitions unchanged: random proposal trigger ($p{=}0.05$), severity fixed at zero, \texttt{own\_transformer\_per\_model} exclusively. Per benchmark, seven arms --- never-adapt, and \{naive, point, strict\} $\times$ candidate size $\in \{512, 2000\}$/class --- give a 21-arm matrix on 30 fresh seeds (4001--4030). The only new variable is the challenger's training size; trigger, probe (32 labels), gate margins, hyperparameters (SVC $C{=}1.0$, $\gamma{=}$\texttt{scale}, PCA dim 8), temporal semantics and sampling policy are unchanged, and the incumbent keeps its 2{,}000/class training set.

One structural property should be stated in advance: at severity zero the 2{,}000-per-class self-contained challenger is drawn from the same pools, by the same sampler and with the same hyperparameters as the incumbent, so it is close to an exchangeable re-draw of the incumbent's own training procedure. Near-zero mean effects at nominal parity are therefore the outcome \emph{expected} of a correct implementation; the control's informative content lies in the direct 512-versus-2{,}000 contrast, which is why the same intervention is repeated under pool-constructed progressive drift in \S\ref{sec:sizematched_drift_method}, where the challenger is no longer an exchangeable copy of the incumbent.

\textbf{Nested candidate batches.} The two size conditions are not independently sampled challengers: at every proposal the draw first produces the \emph{unchanged} 512-per-class batch (bit-identical to the symmetric replication's arms), then extends it with 1{,}488 further draws per class from the same per-trigger RNG stream, so the 512 batch \emph{is} the first 512 rows per class of the 2{,}000 batch (999 proposal-level pairs verified). Pool draws are with replacement, so the extension adds nominal draws, not necessarily 1{,}488 distinct flows or an equal amount of effective information. Each challenger's scaler/PCA/classifier still fit on its own batch only. This is a controlled pool-based construction, not an observed-data arm, and what it matches is \emph{nominal} per-class sample size; \S\ref{sec:sizematched} states what nominal parity does and does not equate.

\textbf{Estimands, families and outcome rules (all frozen).} Primary estimands, in BA points per benchmark: E1 size-matched damage ($\mathrm{naive}_{2000}-\mathrm{never}$), E2 candidate-size effect ($\mathrm{naive}_{2000}-\mathrm{naive}_{512}$), E3/E4 gate value at matched size, E5 (secondary) the gate$\times$size interaction; four Holm-corrected families (F1--F4: 3/3/6/6 contrasts), CI95 for harm, CI90 for equivalence at the $\pm0.5$-point margin, the same guardrails, the seed as inferential unit. Three machine-evaluable outcomes were frozen in advance: PERSISTENCE (material Holm-significant damage at 2{,}000 plus a guardrail-clean gate win), ELIMINATION (equivalence of all three damage contrasts \emph{and} all six gate contrasts, plus no contradicting harmful-future-value signal, keyed to the \emph{sign rate} of committed proposals' five-window future value) and ATTENUATION (anything else). \S\ref{sec:sizematched} applies these rules literally.

\subsubsection{Candidate evidence size under pool-constructed progressive drift}
\label{sec:sizematched_drift_method}

The zero-drift control cannot, by construction, show that candidate evidence matters where recency carries genuinely shifted information. A pre-specified control therefore repeats the nested size intervention under full progressive drift, the regime in which the frozen-transformer harness had reported that size-matching \emph{deepened} harm (Online Resource~1, \S S1.5). Its protocol, outcome rules and analysis script were committed before the drift-time nested draw was implemented, and the implementation was checked bit-for-bit against stored outputs before the confirmatory block ran.

\textbf{Design.} Three full-drift scenarios (PortScan, UNSW-Recon, ToN-IoT; mixing ramp 0$\to$1 over 80 of 100 windows), the same seven arms, on 30 fresh seeds (6001--6030), \texttt{own\_transformer\_per\_model} exclusively, every other constant identical to the symmetric replication. \emph{Nested draw at drift:} at a proposal at window $t$ the draw first produces the 512-per-class batch at the proposal-time severity $\mathrm{sev}(t)$, then continues the same RNG stream to draw the 1{,}488-per-class extension at the same $\mathrm{sev}(t)$, so both sizes sample the same proposal-time mixture and differ only in nominal candidate evidence (prefix-hash and severity equality verified at every coupled proposal). \emph{Coupling scope:} the two always-deploy arms share identical trigger and commit timelines, so their proposals are exactly coupled, whereas gated arms may diverge after the first discordant decision, so gate$\times$size contrasts are seed-paired only.

\textbf{Estimands, families and outcome rules (all frozen).} G1 value of updating ($\mathrm{naive}_{512}-\mathrm{never}$ and $\mathrm{naive}_{2000}-\mathrm{never}$; 6 contrasts), G2 the primary size effect ($\mathrm{naive}_{2000}-\mathrm{naive}_{512}$; 3), G3 gate value at 2{,}000 (6), G4 secondary gate$\times$size interactions (6); Holm within family, the same margin and guardrails. The frozen classification of G2 per scenario is \textsc{size benefit} (Holm-significant, effect $\ge +0.5$), \textsc{size cost} (Holm-significant, $\le -0.5$), \textsc{no material size effect} (CI90 within $\pm0.5$) or otherwise unresolved/sub-material, with a block label \textsc{homogeneous} when all three scenarios share a class and \textsc{heterogeneous} otherwise. Future-value summaries are descriptive only.

\subsection{Promotion policies and the common-harness comparison}
\label{sec:baselines_method}

The primary policy matrix is common to every control: \textbf{never-adapt} (the frozen $h_0$), \textbf{always-deploy} (\emph{naive} in estimand names and tables: commit every confirmed proposal), and the \textbf{point} and \textbf{strict} labeled-probe gates of \S\ref{sec:gates} (strict additionally rejects probe ties, at no additional labels). The formal sequential alternatives --- exact McNemar, the confidence-sequence gates and the assembled VBC-SG with its deferral and lifetime-spending machinery (\S\ref{sec:riskgates}) --- are evaluated as a separate axis under the historical frozen policy, where their guarantees and label costs are priced (\S\ref{sec:risk}).

\textbf{The common-harness comparison.} The common-harness block evaluates published and reference alternatives with self-contained challengers on bit-identical raw streams and source-row-disjoint roles; the exact-feature-disjoint repetition is reported in \S\ref{sec:value_disjoint}. Its protocol was frozen before implementation and, still before any result, placed the \emph{primary} comparison at nominal 2{,}000-per-class evidence parity. Policies: never-adapt, always-deploy, point and strict (anchors); \textbf{ATC} \cite{garg2022atc} and \textbf{DoC} \cite{guillory2021doc}, the two standard published label-free accuracy estimators, used as promotion rules (each model estimates its current-window accuracy from its own confidences plus a 512-row labeled validation sample drawn at training time; zero target-window labels); a \textbf{calibrated soft ensemble} and \textbf{replay 50/50} retraining (standard baselines); and the \texttt{river} reference implementations of \textbf{DDM} and \textbf{ADWIN} \cite{gama2004learning,bifet2007learning} as retraining triggers with always-deploy on fire, run at their registered reference parameters --- the implementations' default DDM thresholds and ADWIN $\delta=0.002$ --- on 8 monitoring labels per window (800 per stream); the monitors were not tuned, so their cells are read as evidence about the evaluated reference configuration, not about the methods in general.

Selection criteria were frozen before any result: closeness to the same decision problem, public reproducibility, dataset compatibility, no information unavailable to competing methods, and checkable implementation fidelity. No faithfully reproducible end-to-end published adaptive-NIDS system was found that matched the same decision problem and information interface (\S\ref{sec:baselines}); none of the evaluated rows is labelled an adaptive-NIDS state-of-the-art method. Six scenarios (three benchmarks $\times$ full and zero drift), 96 arms on 30 fresh seeds (5001--5030): the nine policies at 2{,}000/class plus a secondary 512/class sensitivity block for \{naive, point, strict, ATC, DoC, ensemble\}. Information budgets differ legitimately by method definition and are documented, not equalized (Online Resource~1, \S S10).

\textbf{Families and rules (frozen).} Primary, at 2{,}000/class: PF1 zero-drift loss avoidance and PF2 full-drift benefit retention (each of \{ATC, DoC, ensemble, replay, DDM, ADWIN\} minus naive; 18 contrasts each), PF3 the published estimators minus the point gate (12). Secondary: SF4 the 512/class sensitivity (18) and SF5 method$\times$size interactions (30; seed-paired). Holm within family; per-cell classification \textsc{material gain} / \textsc{material cost} (Holm-significant, $|\Delta| \ge 0.5$), \textsc{compatible} (CI90 within $\pm0.5$) or \textsc{unresolved}; anchor-versus-anchor contrasts are descriptive here so that the hypotheses of \S\ref{sec:sizematched_method}--\S\ref{sec:sizematched_drift_method} are not double-tested. Four pre-specified statements (S1--S4, Online Resource~1, \S S10) are evaluated literally.

\subsection{Exact-cleaned-feature-disjoint integrity sensitivity}
\label{sec:value_disjoint_method}

The historical confirmatory design separated roles by source row: the 50/30/20 splitter prevents reuse of a source-row index, but an audit of the cleaned data found exact cleaned raw feature vectors in more than one role. Exact identity was defined on the complete cleaned raw vector after the experiment's numeric coercion and NaN/infinity handling, as canonical float64 bytes with signed zero normalized; no rounding, PCA, scaling or approximate matching entered the key, SHA-256 grouping was confirmed by vector equality, and the key used features only (labels, class, reference/current membership and row identity excluded). Across the three scenarios, 13.63\%, 36.41\% and 11.23\% of rows were beyond the first member of an exact-$X$ group; UNSW-NB15 contained 403 conflicting-label groups involving 1{,}828 rows. Every original row and label was retained.

One final protocol, committed before implementation or results, replaced only role assignment: for each seed, all rows sharing exact $X$ are assigned to one role by a deterministic seed-dependent constrained greedy algorithm that targets the pre-specified stratum fractions for reference/current $\times$ benign/attack, preserving multiplicity, contradictory labels and every source row while guaranteeing $X(\mathrm{window})\cap X(\mathrm{train})=X(\mathrm{window})\cap X(\mathrm{probe})=X(\mathrm{train})\cap X(\mathrm{probe})=\varnothing$. Candidate sampling remains with replacement within its training role; the 512/class batch remains the prefix of the 2{,}000/class batch, and both use the same proposal-time mixture. All role audits passed, with zero cross-role exact-$X$ overlap and maximum stratum-fraction deviation 0.0058 percentage points.

The sensitivity repeated the 21 full-drift arms of the size control under drift (block B2 below) on fresh seeds 7001--7030 and the 96 arms of the common-harness comparison (block B1) on fresh seeds 8001--8030. B2 froze magnitude-aware classes \textsc{robust homogeneous size benefit}, \textsc{partial robustness}, \textsc{no material size effect}, \textsc{size cost} and \textsc{heterogeneous}; B1 reused its original families and per-cell rules and froze \textsc{policy conclusions robust}, \textsc{partially robust} or \textsc{materially changed}. Smoke seeds and confirmatory blocks were disjoint; grouping used features only and was fixed before any result was inspected, and no parameter changed after execution began. The exact protocol, configs, implementation tests, audit and complete results are in the artifact and Online Resource~1, \S S11.

\subsection{Outcomes and inference}
\label{sec:outcomes}

Per window we record balanced accuracy (BA) and attack-class F1; per stream we average over the 100 windows, and per condition over seeds. BA is the primary endpoint and measures the mean effect; attack recall and false-positive rate are pre-specified non-inferiority guardrails that gate safety \emph{language}, not the verdicts; harmful-proposal summaries are descriptive within-trajectory fractions over seed-clustered commits; probe-level guarantees are not future-deployment guarantees. We report \textbf{gain vs. no-adaptation} (the frozen $h_0$ baseline), \textbf{committed adaptations} and \textbf{labels used}; BA-point differences are not converted into misclassified-flow counts, which would require a prevalence model the balanced pool windows do not supply. The per-stream \textbf{policy oracle} $\mathrm{BA}_{\text{oracle}}=\mathbb{E}_{\text{seed}}\max(\mathrm{BA}_{\text{no-adapt}},\mathrm{BA}_{\text{detector}})$ bounds the value of perfect per-trigger decisions from below, the \textbf{regret} of always-deploy is $\mathrm{BA}_{\text{oracle}}-\mathrm{BA}_{\text{naive}}$, and the \textbf{harm frequency} is the fraction of streams on which adapting scored below no-adaptation. All confirmatory comparisons are \textbf{truly paired} on bit-identical pre-generated streams, with cluster (per-seed) bootstrap 95\% confidence intervals; the initial exploratory study uses seed-matched bootstrap CIs (conservative rather than fully paired).

Generative-AI assistance was used in manuscript editing and in implementation and review of analysis scripts under author supervision; the tools, scope and human-accountability statement are disclosed in the Declarations.

\textbf{Evidence stages.} The study combines an exploratory discovery stage, the frozen-policy replication, two comparability controls (transformer ownership; zero-drift candidate size), the source-row-disjoint full-drift size control and common-harness comparison, their final exact-feature-disjoint sensitivity, supporting pre-specified analyses, and chronological boundary evaluations. The final claims use the exact-feature-disjoint result to qualify the earlier blocks; the remaining blocks test mechanisms, costs or boundaries. The full evidence hierarchy, per-block configuration, seeds and provenance are catalogued in Online Resource~1, \S S0.

\section{Results}

Results follow the pipeline of Fig.~\ref{fig:pipeline}: the historical configuration that exposed the promotion problem (\S\ref{sec:historical}), candidate construction (\S\ref{sec:symmetric}), candidate evidence at zero drift (\S\ref{sec:sizematched}) and under pool-constructed progressive drift (\S\ref{sec:sizematched_drift}), the conditional value of validation (\S\ref{sec:gatevalue}), the common-harness comparison (\S\ref{sec:baselines}), the exact-feature-disjoint sensitivity (\S\ref{sec:value_disjoint}), and mechanism, formal and boundary evidence, including the chronological replays (\S\ref{sec:boundaries}). Table~\ref{tab:synthesis} maps the source-row-disjoint evidence; Table~\ref{tab:value_disjoint_main} supplies the final sensitivity that governs the size-effect conclusion.

\input{tables_springer/table_synthesis.tex}

\subsection{The historical frozen configuration: where the promotion problem appeared}
\label{sec:historical}

Under the study's historical configuration --- a frozen, incumbent-owned scaler and PCA reused by every challenger, and 512-per-class challengers against a 2{,}000-per-class incumbent --- a replication on the hardened harness (seeds 104--133; full table in Online Resource~1, \S S2.11) established the reference behaviour that motivated what follows: always-deploy updating is net-harmful in the harm regime ($-1.64$ [$-2.72$, $-0.64$] balanced-accuracy points on ToN-IoT) and the 32-label commit gate converts it to a net benefit ($+0.79$; $+2.43$ above always-deploy) while preserving the benefit and marginal regimes, under both a classical and a quantum-kernel detector. The same configuration makes always-deploy net-harmful in all three benchmarks once the incumbent stays healthy, under mild drift ($-0.46$, $-0.15$, $-0.65$) and under zero drift with random proposals ($-2.76$, $-0.75$, $-4.75$; Online Resource~1, \S S2.12), and a statistical commit rule that refuses to commit without evidence recovers the zero-drift loss almost entirely.

\textbf{Absolute harm magnitudes in this configuration vary between seed blocks.} The frozen configuration above was executed on three independent 30-seed blocks with identical data, flags, trigger semantics, candidate sampler and learner (the released code reproduces this block's per-seed outputs bit-for-bit), and the absolute always-deploy harm on ToN-IoT full drift differs materially between them: $-1.64$ [$-2.72$, $-0.64$] here (seeds 104--133), $-1.97$ [$-4.00$, $-0.14$] for the always-deploy anchor of the registered budget frontier (seeds 501--530; \texttt{frontier\_anchors.csv} in the artifact) and $-4.95$ [$-7.15$, $-2.86$] in the frozen arm of the symmetric replication (seeds 3001--3030; Table~\ref{tab:symmetric_pipeline}); the zero-drift frozen cells show the same pattern with overlapping intervals ($-2.76$ here against $-4.09$ on PortScan in Table~\ref{tab:synthesis}). The per-seed distribution is heavy-tailed: in the symmetric replication's frozen arm the median stream loses $2.4$ points while seven of thirty streams lose more than ten, because a committed challenger occasionally collapses, so a block's mean depends on how many such streams it draws.

Any single absolute frozen-harm magnitude is therefore quoted only as a descriptive block-level value; the comparative conclusions rest on paired within-block contrasts (the same streams under both policies), and the self-contained size-matched and common-harness results of \S\ref{sec:sizematched}--\S\ref{sec:baselines} do not use these historical absolute values.

Read on its own, this block says ``promotion harms a healthy incumbent, and a gate rescues it''. The rest of the paper shows why that reading is a statement about the configuration: who owns the preprocessing, and how much evidence the challenger sees, drive the harm and most of the gate's measured value. The frozen-policy numbers are retained as diagnostic evidence only (full grids: Online Resource~1, \S S2.11--S2.12).

\subsection{Candidate construction: preprocessing ownership changes the apparent value of promotion}
\label{sec:symmetric}

Table~\ref{tab:symmetric_pipeline} reports the replication of \S\ref{sec:symmetric_method}; its pre-specified verdict was Scenario~A --- \emph{persistence with a shifted locus}: preprocessing asymmetry amplified harmful promotion, while a residual zero-drift harm survived for self-contained challengers. All of its challengers train on 512 flows per class against a 2{,}000/class incumbent, the asymmetry \S\ref{sec:sizematched} then isolates; the gate-value family is read in \S\ref{sec:gatevalue}.

\textbf{(1) The mean full-drift harm does not persist under self-contained pipelines.} With each challenger owning its preprocessing, always-deploy is \emph{beneficial} in all three full-drift regimes: PortScan $+7.21$ [4.89, 9.52], UNSW-Recon $+2.55$ [2.34, 2.75], and --- the regime where the harm was originally reported --- ToN-IoT $+1.03$ [0.55, 1.53], all Holm-significant.

\textbf{(2) Preprocessing ownership is a major amplifier --- not a complete explanation of candidate-promotion risk.} The ownership interaction $\mathrm{BA}_{\text{naive,own}} - \mathrm{BA}_{\text{naive,frozen}}$, on bit-identical raw streams and candidate batches, is positive everywhere it was harmful: $+5.98$ [3.82, 8.26] on ToN-IoT full, $+0.90$ and $+1.15$ on PortScan and UNSW full, and $+5.49$/$+2.36$ on the ToN/PortScan zero-drift streams. The frozen configuration had been handicapping every challenger, most severely where the harm was reported: the original full-drift harm was configuration-dependent and strongly amplified by incumbent-owned preprocessing. Freezing a representation can be a legitimate production policy; the issue is generalizing results obtained under that treatment beyond it.

\textbf{(3) But ownership does not explain all of the risk: zero-drift promotion harm persists at the historical candidate size.} With no drift at all, self-contained 512-per-class challengers still hurt: naive$-$never is $-1.74$ [$-2.35$, $-0.96$] on PortScan and $-0.65$ [$-0.77$, $-0.55$] on UNSW-Recon --- both beyond the preregistered $\pm0.5$-point materiality margin, Holm-significant --- and $-0.38$ [$-0.53$, $-0.25$] on ToN-IoT, sub-material but significant and \emph{not} equivalent to zero (its CI90 is not inside $\pm0.5$). These challengers were also trained on a quarter of the incumbent's evidence, and \S\ref{sec:sizematched} shows that asymmetry, not re-estimation per se, accounts for the mean effect.

\input{tables_springer/table_symmetric_pipeline.tex}

\textbf{Security guardrails and future-negative signs.} Every BA-winning self-contained gate cell passes both non-inferiority margins except one: on zero-drift UNSW-Recon, strict validation improves balanced accuracy and false-positive rate, but its attack-recall lower bound crosses the 1-point margin ($\Delta$recall $-0.99$, bound $-1.26$), and that cell is not described as a security improvement anywhere in this paper; the FPR point estimate improves in each of the other winning cells (up to $-3.33$ points for strict on zero-drift PortScan), while several non-winning full-drift gate cells fail the FPR margin (Online Resource~1, \S S7). Under self-contained always-deploy updating at 512/class, the descriptive fraction of committed proposals with future-negative five-window sign is 65\%/61\%/46\% on the zero-drift streams (42\% on ToN full); commits cluster within seeds and support no population-prevalence or deployment-probability inference. The quantum monitors, the VBC-SG frontier and the mild-drift matrix were not re-run under self-contained pipelines; their evidence remains scoped to the frozen policy (\S7).

\subsection{Candidate evidence at zero drift: nominal candidate-evidence size accounts for the residual zero-drift mean harm}
\label{sec:sizematched}

Table~\ref{tab:size_matched} reports the control of \S\ref{sec:sizematched_method}; the gate family (F3) is read in \S\ref{sec:gatevalue}.

\textbf{(1) The 512 harm replicates; matched-size means are compatible with the margin.} On thirty fresh seeds, 512-per-class challengers reproduce the residual harm almost exactly (naive$_{512}-$never: $-1.70$ [$-2.05$, $-1.36$] PortScan, $-0.65$ [$-0.77$, $-0.55$] UNSW, $-0.24$ [$-0.39$, $-0.10$] ToN; descriptive). Give the same challengers the incumbent's 2{,}000 flows per class, as a pure extension of the nested batch, and naive$_{2000}-$never is $+0.19$ [$-0.15$, $+0.55$], $-0.02$ [$-0.10$, $+0.06$] and $-0.01$ [$-0.05$, $+0.04$], all compatible with the $\pm0.5$-point margin (CI90 inside it) and none approaching Holm significance. PortScan is margin-sensitive at $\pm0.2$: its CI90 upper bound is $0.494$, only $0.006$ points inside the margin, so that cell is margin-dependent and does not demonstrate absence of an effect; UNSW-NB15 and ToN-IoT pass at $\pm0.2$, and all three at $\pm1.0$.

\textbf{(2) Candidate size is the driver, and it is Holm-significant everywhere.} The paired size effect naive$_{2000}-$naive$_{512}$ is $+1.89$ [$+1.54$, $+2.23$] on PortScan, $+0.63$ [$+0.53$, $+0.73$] on UNSW and $+0.23$ [$+0.10$, $+0.38$] on ToN --- essentially the mirror image of the 512 harm, all three Holm-significant. Increasing candidate evidence made each mean effect compatible with the margin in the evaluated zero-drift control; it did not establish absence of an effect, nor equivalence in every respect a proposal can differ (Online Resource~1, \S S8).

\textbf{What nominal sample-size parity does and does not match.} The matched condition equates the challenger's and the incumbent's \emph{nominal} per-class training size: 2{,}000 rows per class against 2{,}000 rows per class. It does not establish equality in effective sample size, temporal coverage, temporal diversity, subtype support, duplication, label quality, prevalence or information content. Pool draws are with replacement, so nominal rows can contain duplicates and should not be read as effective sample sizes; the candidate batch is a controlled balanced sample from the current-regime pool, not a time-ordered slice of observed traffic; and neither condition equates the temporal span the two training sets cover. Within these controlled conditions the experiment isolates the nominal row-count factor, and that factor accounts for the residual mean harm. Nominal parity is not information parity: evidence comparability in deployment remains an operational property to be audited, and \S7 records the corresponding scope limits.

\input{tables_springer/table_size_matched.tex}

\textbf{Pre-specified outcome: ATTENUATION --- a property of the rule, not evidence of residual harm.} The frozen rules classify this result mechanically. PERSISTENCE fails because there is no material mean damage at the matched size (no F1 contrast is $\le -0.5$ with CI95 below zero). ELIMINATION fails on exactly one criterion: the harmful-future-value check (E3) keys on the \emph{sign rate} of committed proposals' five-window value, and at the matched size that rate is $\approx$48--52\% in two benchmarks, so the outcome is ATTENUATION. When the mean future value of committed challengers is approximately zero --- which is what near-exchangeable challengers make expected (\S\ref{sec:sizematched_method}) --- negative and positive signs occur at close to equal frequency, so a sign-rate threshold of this form is structurally unable to certify elimination in exactly the situation it was registered to detect. The label is preserved mechanically and should be read as a property of the rule, not as evidence of directional residual harm; the same logs show a small mean five-window future value of committed 2{,}000-challengers ($-0.06$ to $+0.03$ points, against $-0.13$ to $-1.03$ at 512), and a near-50\% negative-sign rate must be read as proposal-level variability within trajectories, not a deployment-risk probability. The formal classification and the margin-based mean comparison are distinct; neither demonstrates absence of an effect, and the full-drift control of the next subsection does not retroactively change this classification.

\textbf{Guardrails at the matched size.} All six matched-size gate cells pass both non-inferiority margins vs.\ naive$_{2000}$, while at 512 the zero-drift UNSW strict cell again fails recall non-inferiority, replicating the earlier trade-off on fresh seeds (Online Resource~1, \S S8). There is no security trade-off at the matched size; there is also no BA gain, and the absence of a trade-off is not evidence of utility.

\subsection{Candidate evidence under pool-constructed progressive drift: the source-row-disjoint result}
\label{sec:sizematched_drift}

The zero-drift null was expected of any correct implementation, because a 2{,}000-per-class challenger drawn from the incumbent's own pools at severity zero is close to an exchangeable copy of the incumbent. The full-drift control of \S\ref{sec:sizematched_drift_method} answers in the only regime where candidate evidence is a non-trivial hypothesis: the challenger now samples a drifted mixture the incumbent has never seen. Table~\ref{tab:size_matched_drift} reports it (21 arms, seeds 6001--6030; nesting and severity audits verified at all 707 coupled proposals).

\textbf{(1) Source-row-disjoint primary result.} In the source-row-disjoint block, increasing the candidate from 512 to 2{,}000 rows per class changes BA by $+0.82$ [$0.67$, $0.98$] on PortScan, $+1.66$ [$1.51$, $1.81$] on UNSW-Recon and $+1.00$ [$0.60$, $1.39$] on ToN-IoT --- all Holm-significant and above the $+0.5$-point materiality margin. Its frozen classification is \textsc{homogeneous-size benefit}. The measurements remain valid for that design, but the design is source-row-disjoint rather than exact-feature-disjoint; the exact-feature-disjoint repetition of \S\ref{sec:value_disjoint} (Table~\ref{tab:value_disjoint_main}) is the governing estimate --- $+0.53$, $+1.67$ and $+0.38$ --- and narrows the conclusion.

\textbf{(2) What this establishes, and what it does not.} Within this pool-based design, increasing nominal challenger evidence improves promotion under a constructed drift trajectory as well as under the zero-drift control, and the frozen-transformer precedent in which size-matching deepened full-drift harm does not reproduce with self-contained pipelines. It does \emph{not} establish effective information parity, a universal monotonic benefit from more data, causal generality beyond the nested sampling design, or an effect free of exact-value exposure; the exact-feature-disjoint estimate in \S\ref{sec:value_disjoint} supersedes the homogeneity claim.

\input{tables_springer/table_size_matched_drift.tex}

\textbf{(3) Validation at the matched size under drift.} Point and strict validation add no positive average value at 2{,}000/class in any of the six G3 contrasts (point $+0.15$, $-0.06$, $+0.01$ over always-deploy, none Holm-significant; strict $-0.00$, $-0.34$, $-0.14$, the UNSW-Recon cell a \emph{resolved cost}, $-0.34$ [$-0.53$, $-0.15$]), and the secondary gate$\times$size interactions are $\le 0$ in every point estimate. Every matched-size point cell passes both guardrails; strict at 2{,}000 fails the FPR margin on UNSW-Recon ($\Delta$FPR $+0.86$), so no safety language attaches to that cell, and at 512 strict fails the recall margin on PortScan and ToN-IoT. The only resolved gate gain of this block is the descriptive point gate on ToN-IoT at 512 ($+0.46$ [$0.09$, $0.88$] over naive$_{512}$), consistent with validation earning its keep where the challenger is evidence-disadvantaged (Online Resource~1, \S S9).

\subsection{Validation has conditional average value}
\label{sec:gatevalue}

The gate families of the three comparability controls, read together, locate where validation earns its keep --- and where it measurably does not.

\textbf{Under a nominal-evidence disadvantage, validation retains material value.} At self-contained 512/class in the symmetric replication (seeds 3001--3030), point and strict validation improve over always-deploy in all six zero-drift comparisons (Holm-significant), recovering much or nearly all of the loss depending on the cell: strict gains $+1.68$ [1.23, 2.13] over always-deploy on PortScan, $+0.51$ [0.38, 0.64] on UNSW, $+0.34$ [0.20, 0.49] on ToN (point $+0.75/+0.19/+0.11$). ``Recover'' means improvement \emph{over always-deploy}: strict returns to within $0.06$--$0.14$ points of never-adapt, the point gate still sits $0.27$--$0.99$ points below it, and no zero-drift cell converts to a net benefit over never adapting, nor should it where there is nothing to gain. At full drift the point gate still filters harmful candidates on ToN-IoT ($+0.64$ [0.26, 1.06]), adds nothing material on PortScan, and has a small resolved \emph{cost} on UNSW-Recon ($-0.21$ [$-0.36$, $-0.06$]): where updating is already clearly beneficial, validation's marginal value can be zero or slightly negative (Online Resource~1, \S S7).

\textbf{At nominal evidence parity, validation has nothing left to recover --- at zero drift and under drift alike.} All six F3 gate contrasts at the matched size under zero drift are tiny ($+0.01$ to $+0.13$ points), none Holm-significant and every one CI90-equivalent to always-deploy within $\pm0.5$; the secondary F4 interactions are uniformly negative and Holm-significant (strict on PortScan $-1.54$ [$-1.89$, $-1.20$]): the gates' value at 512 was, to the extent the interaction measures it, compensation for the challenger's evidence disadvantage (Online Resource~1, \S S8). Under full drift the G3 family repeats the pattern with fresh seeds: 0/6 positive gate effects at 2{,}000/class and a resolved strict-gate cost on UNSW-Recon (\S\ref{sec:sizematched_drift}). No average benefit under these controls is not the same as no value: it does not rule out value against individual harmful candidates, tail risk, temporally uncertain streams, or configurations not evaluated here.

\textbf{Evidence versus validation: what each nominal label buys.} Table~\ref{tab:evidence_validation_tradeoff} compares, per benchmark, the 512-per-class policies against always-deploy at nominal 2{,}000-per-class parity, from the size-matched control (seeds 4001--4030), so its 512-side gate values re-estimate on fresh seeds the contrasts quoted above from seeds 3001--3030. In nominal adjudication counts, moving from 512 to 2{,}000 samples per class requires $2\times1{,}488 = 2{,}976$ additional candidate-training labels per proposal, whereas the point/strict decision uses a 32-label probe. At 512 the strict gate recovers most of the mean loss (PortScan: naive$_{512}$ $-1.70$ vs strict$_{512}$ $-0.03$ against never, descriptive); at 2{,}000 the mean effects are compatible with the $\pm0.5$-point margin without a probe. This establishes no economic dominance in either direction: labels are counted, not priced, and the inspected-flow cost of acquiring balanced attack labels is not modeled.

\input{tables_springer/table_evidence_validation_tradeoff.tex}

Under the historical frozen policy the same conditionality appears from the other side: with a healthy incumbent and asymmetric challengers, point/strict and the risk-controlled rules recover most or all of the zero-drift loss (\S\ref{sec:v2}), while the chronological replays (\S\ref{sec:boundaries}) show what conservative validation forgoes when the incumbent has collapsed.

\subsection{Source-row-disjoint common-harness comparison with published and reference baselines}
\label{sec:baselines}

Table~\ref{tab:common_harness} reports the source-row-disjoint comparison of \S\ref{sec:baselines_method} at its primary condition: self-contained challengers at nominal 2{,}000-per-class evidence parity, 96 arms on seeds 5001--5030, every policy on bit-identical streams. Recent end-to-end adaptive NIDS have different decision units and label/update interfaces (\S2; Online Resource~1, \S S11), so transplanting them would not reproduce their published decision problem; cross-policy information budgets differ by definition and are reported, never equalized (Online Resource~1, \S S10). The exact-feature-disjoint B1 sensitivity in \S\ref{sec:value_disjoint} governs the final policy interpretation.

\input{tables_springer/table_common_harness.tex}

\textbf{Zero drift at parity: nothing left to rescue.} With 2{,}000-per-class challengers always-deploy is already approximately neutral relative to never-adapt ($+0.25$, $+0.07$, $+0.02$; descriptive), while the 512-per-class anchor replicates the residual harm on a third fresh seed block ($-1.66$, $-0.58$, $-0.35$). No policy attains the pre-specified statement S1 (``avoids the zero-drift loss''): every PF1 cell is \textsc{compatible} or \textsc{unresolved} except the calibrated ensemble on PortScan (\textsc{material gain}, $+1.46$); the zero-drift problem that label-free alternatives were previously priced against disappears once the challenger is comparably evidenced.

\textbf{Full drift at parity: two compatible alternatives, four that pay as configured.} \textbf{ATC} is \textsc{compatible} with always-deploy on UNSW-Recon ($-0.22$) and ToN-IoT ($+0.10$) and \textsc{unresolved} on PortScan ($-0.65$); the \textbf{calibrated ensemble} is \textsc{compatible} in all three ($-0.09$, $-0.42$, $-0.23$). \textbf{DoC} pays a \textsc{material cost} on PortScan ($-2.59$) and UNSW-Recon ($-0.59$) and is \textsc{compatible} on ToN-IoT ($+0.21$); \textbf{replay} ($-0.89$, $-2.07$, $-1.22$), \textbf{river-DDM} ($-0.78$, $-2.67$, $-1.81$) and \textbf{river-ADWIN} ($-8.73$, $-4.06$, $-2.50$) are \textsc{material cost} in every full-drift scenario, the two monitors while also consuming 800 monitoring labels per stream (ADWIN's cost is largely under-triggering: at its registered reference parameters it barely fires, so its arm stays close to never-adapt; neither monitor was tuned, so these cells characterize the evaluated reference configuration, not the methods in general).

Statement S2 (``pays for it at full drift'') therefore holds for DoC, replay, DDM and ADWIN and for neither ATC nor the ensemble. ATC is the strongest published zero-target-label competitor; the calibrated ensemble is the strongest standard label-free baseline (it commits every trigger and cannot decline an update, and it fails recall non-inferiority on PortScan). Compatibility here is the pre-specified CI90-within-$\pm0.5$ criterion, not a demonstration of equality.

\textbf{Against the point gate.} ATC is \textsc{compatible} with the point gate on five of six scenarios and \textsc{unresolved} on PortScan full drift ($-0.73$), so statement S3 (``matches the point gate'') is narrowly not attained; DoC is a \textsc{material cost} relative to the gate on PortScan ($-2.67$) and UNSW-Recon ($-0.58$) full drift and \textsc{compatible} elsewhere. The exploratory-harness observation that DoC beat the point gate in the harm regime (Online Resource~1, \S S1.5) does not reproduce under the final harness at parity (ToN-IoT full drift: DoC$-$point $+0.01$, \textsc{compatible}). Neither point nor strict validation dominates (descriptive full-drift differences to always-deploy at 2{,}000: $+0.08$/$-0.00$/$+0.20$ and $-0.16$/$-0.23$/$-0.22$, the UNSW strict cell resolved negative). Many full-drift cells of the label-free and monitor policies fail the FPR guardrail (ADWIN $\Delta$FPR up to $+9.5$ points; Online Resource~1, \S S10), and no safety language attaches to them.

\textbf{The 512/class sensitivity: method ordering depends on the candidate generator.} With evidence-disadvantaged 512-per-class challengers every label-free alternative gains at zero drift (the ensemble $+2.93$/$+0.20$/$+0.36$, DoC $+1.93$/$+0.40$/$+0.44$, ATC $+1.65$/$+0.25$/$+0.26$ over naive$_{512}$), because naive$_{512}$ is itself harmful there, and ATC/DoC surrender full-drift benefit on PortScan ($-1.82$, $-3.63$) while gaining on ToN-IoT ($+0.48$, $+0.78$). All fifteen Holm-significant method$\times$size interactions (SF5) are negative, and the pre-specified ordering-change rule S4 fires for ATC (PortScan zero drift: \textsc{material gain} at 512, \textsc{compatible} at 2{,}000) and for the calibrated ensemble (PortScan and ToN-IoT full drift), not for point, strict or DoC. The value of every evaluated safeguard, labeled gates and label-free alternatives alike, concentrates where candidate evidence is asymmetric. This is the paper's central methodological point in its sharpest form: even conclusions about \emph{which update policy appears preferable} depend on the comparability of the candidate generator, so a policy ranking obtained with evidence-disadvantaged challengers need not survive at parity.

The historical policy blocks --- the frozen-policy comparison, the frozen zero-drift control, the exploratory-harness estimates and the lifetime-budgeted frontier --- are retained in Online Resource~1, \S S2.12 with their provenance and evidence tiers; where the same method and question are evaluated here, the common-harness evidence supersedes them, and cross-block comparisons remain descriptive. The comparison is a map of evaluated accuracy/label/update trade-offs, not a state-of-the-art ranking.

\subsection{Final exact-feature-disjoint sensitivity: positive size effect, partial policy robustness}
\label{sec:value_disjoint}

The audit verified the exposure rather than assuming it (\S\ref{sec:value_disjoint_method}): under the source-row splitter, exact-$X$ groups span window, train and probe roles, and candidate/future exposure differs with candidate size (Online Resource~1, \S S11). These diagnostics establish exposure, not its causal effect.

\input{tables_springer/table_value_disjoint_main.tex}

\textbf{B2: the direction survives, homogeneity does not.} Under exact-feature-disjoint roles, the paired size effect remains positive and resolved in all three benchmarks: $+0.53$ [0.27, 0.79] on PortScan, $+1.67$ [1.47, 1.89] on UNSW-Recon and $+0.38$ [0.21, 0.57] on ToN-IoT, all Holm-resolved (Table~\ref{tab:value_disjoint_main}). PortScan and UNSW meet the pre-specified $+0.5$-point materiality rule on the registered mean estimand, although the PortScan seed distribution remains heterogeneous around that margin; ToN-IoT is resolved but sub-material. The pre-specified verdict is \textsc{partial robustness}; the source-row block's \textsc{homogeneous-size benefit} classification is therefore not retained as the headline. Point gating adds no resolved benefit at parity and strict gating has one resolved sub-material cost on UNSW-Recon. The between-block sensitivity-minus-historical contrast on the size effect is compatible on UNSW ($+0.01$), unresolved on PortScan ($-0.29$) and materially attenuated on ToN-IoT ($-0.62$, Holm $p=0.0149$); the pre-specified rule therefore does not classify the original size effect as materially inflated overall, and these independent-block changes are not causal estimates of duplicate exposure.

\textbf{The absolute value of updating also moved between blocks.} Always-deploy at 2{,}000/class retains a positive mean effect over never-adapt in all three benchmarks ($+7.10$ [5.26, 8.96], $+3.90$ [3.56, 4.26] and $+1.06$ [0.37, 1.92]; ToN-IoT Holm $p=0.017$), although the ToN-IoT seed distribution is highly heterogeneous and its median effect is near zero; at 512/class the corresponding mean gains are $+6.57$, $+2.23$ and $+0.68$, the last no longer statistically resolved (CI95 [0.01, 1.52], Holm $p=0.074$). The exact-feature-disjoint block therefore also yielded smaller absolute always-deploy gains in PortScan and ToN-IoT than the earlier independent block (Table~\ref{tab:size_matched_drift}: $+9.72$ and $+2.44$ at 2{,}000/class, $+8.90$ and $+1.44$ at 512/class), whereas UNSW-NB15 was similar ($+3.79$ and $+2.13$). Because these are independent seed blocks, this comparison is descriptive rather than a causal attribution to duplicate exposure.

\textbf{Security metrics behind the size effect.} In the same sealed cells the balanced-accuracy advantage of 2{,}000/class over 512/class always-deploy is driven primarily by a lower false-positive rate --- $5.44\to4.70$ (PortScan), $31.20\to27.51$ (UNSW-Recon) and $28.39\to27.81$ (ToN-IoT) percentage points --- while attack recall is approximately stable ($89.82\to90.13$ and $97.16\to97.34$) and slightly lower on UNSW-Recon ($97.86\to97.51$; Table~\ref{tab:value_disjoint_main}). The size benefit should therefore be read as fewer false alarms at approximately unchanged attack recall, not as improved attack detection; a lower false-positive rate is not called better wherever a recall guardrail is crossed. The absolute false-positive rates of roughly 27--31\% in the UNSW-Recon and ToN-IoT full-drift cells are not deployment-grade: this study evaluates promotion behaviour under the chosen benchmark pipeline, not a production-ready detector.

\textbf{B1: policy interpretation is partially robust.} Four of six pre-specified robustness predicates hold, with no direct historical material-gain/material-cost reversal. No evaluated policy globally dominates, policy ordering remains candidate-size-dependent, ATC remains compatible with the point gate in five of six scenarios, and material costs remain among alternatives. The stronger source-row retention claims do not survive intact: at full drift ATC is unresolved on PortScan/UNSW and compatible on ToN-IoT, and the calibrated ensemble is unresolved/material cost/compatible; DoC costs materially on PortScan and UNSW, replay and DDM on UNSW, and ADWIN in all three, always at the registered reference parameters for DDM/ADWIN. Point/strict anchors add no positive material effect at parity. The pre-specified B1 verdict is \textsc{partially robust}: validation and policy rankings remain conditional on construction and evidence, and no policy is globally dominant.

The thesis that survives the sensitivity is therefore narrower than the source-row blocks suggested: nominal evidence has a positive resolved effect in all three exact-feature-disjoint benchmarks, but material benefit is benchmark-dependent; policy ordering remains size-dependent; and validation remains conditional. No causal claim is made about exact duplicates, and no generalization extends beyond exact duplicate-value exposure.

\subsection{Mechanism, formal instruments and external boundaries}
\label{sec:v2}
\label{sec:risk}
\label{sec:boundaries}

The remaining blocks support the mechanism behind the results above, price the formal and operational cost of validating, and delimit external validity; Table~\ref{tab:evidence_map} states each block's question, main finding and scope, and the full grids are in Online Resource~1. Except where noted, these blocks are evidence under the historical frozen-transformer configuration at 512/class, and their numbers should not be assumed to transfer to self-contained, size-matched pipelines.

\begin{table*}[t]
\centering
\caption{Supporting evidence blocks. Each is detailed in Online Resource~1 at the section shown; scopes are stated in \S7.}
\label{tab:evidence_map}
\footnotesize
\resizebox{\ifdim\width>\linewidth \linewidth\else\width\fi}{!}{%
\begin{tabular}{p{0.15\linewidth} p{0.23\linewidth} p{0.30\linewidth} p{0.20\linewidth}}
\toprule
Extension & Question & Main result & Scope / limitation \\
\midrule
Observed-data (\S S2.7) & Survives without simulator privileges? & Harm-regime rescue reproduces leakage-free ($+$4.05/$+$4.83 above always-deploy) & Frozen, 512/class; UNSW below no-adapt at 64-flow windows \\
Classifier / generator (\S S1.6, S2.12) & Is the harm SVC-specific? & Sign general at 512; size-matching clears robust learners & Magnitude pipeline-dependent; frozen transformer \\
Mild drift (\S S2.12) & Is harm dataset-specific? & Healthy incumbents make always-deploy harmful in all three & Frozen, 512/class \\
Quantum monitor (\S S1.2) & Does a more expressive detector change conclusions? & No & ${\sim}114\times$ simulated cost; invariance check only \\
VBC-SG frontier (\S S2.12) & What do formal guarantees cost? & Non-vacuous from $b{=}256$; abstention/labels priced & Frozen; not necessary under the size-matched controls \\
Chronological replays (\S S2.6) & External time-ordered boundary? & No net harm observed in any of 13 replays (7 in Table~\ref{tab:chronological_q1}, 6 earlier); gates pay a premium on collapsed incumbents & No ToN-IoT analogue; no prevalence estimate \\
Prevalence / acquisition (\S S2.3) & What does finding attack labels cost? & ${\approx}1/\pi$ inspected flows; alert-enrichment cuts it 5--8$\times$ & Not end-to-end; starved trigger is the upstream limit \\
\bottomrule
\end{tabular}}
\end{table*}

\begin{figure*}
\centering
\includegraphics[width=\textwidth]{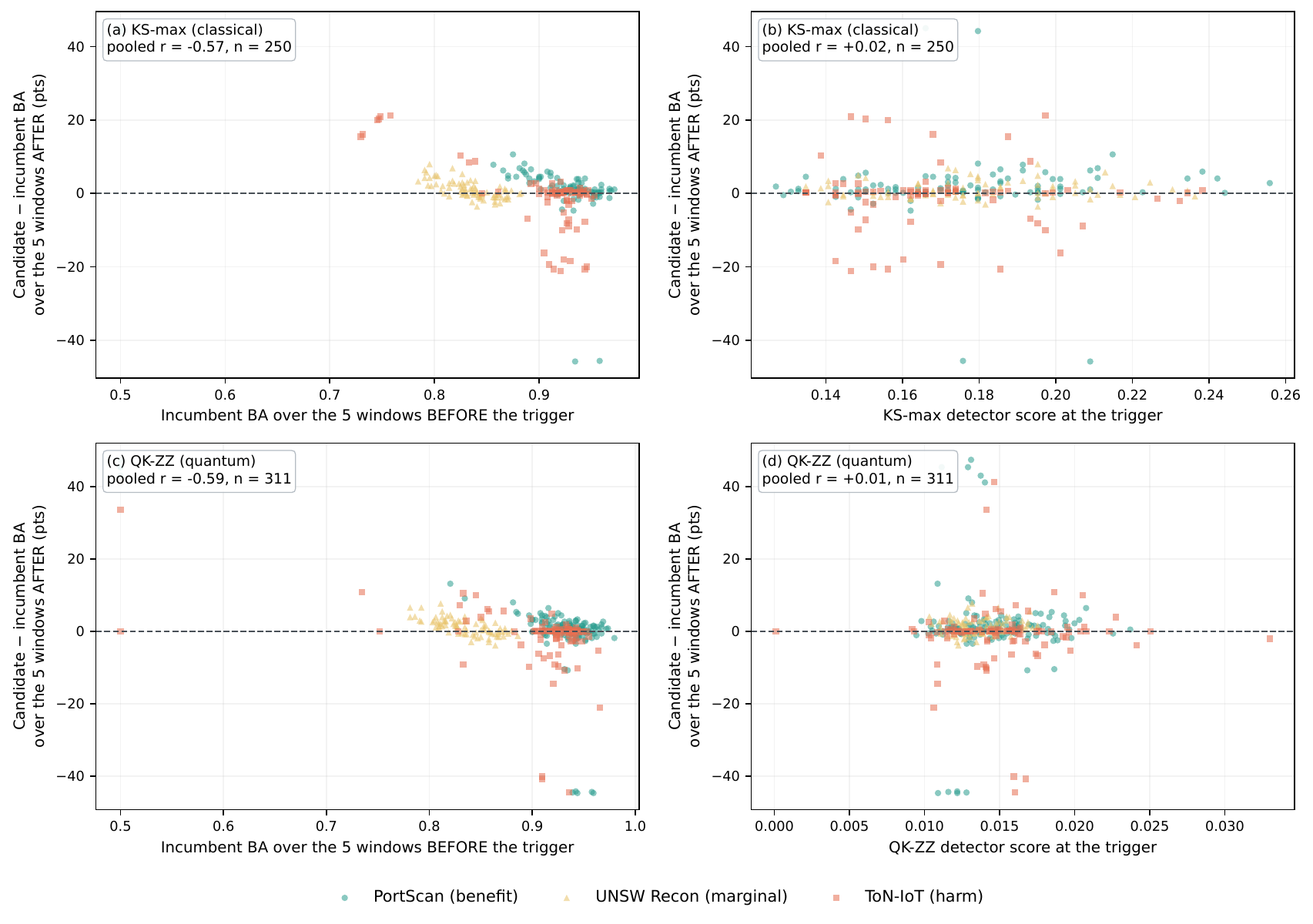}
\caption{Per-trigger test on the hardened harness (seeds 104--133; 250 KS-max triggers): incumbent balanced accuracy over the five windows before the trigger (left) and detector score at the trigger (right), each against the candidate-minus-incumbent balanced accuracy over the five following windows, for the KS-max (top) and QK-ZZ (bottom) detectors; colours and marker shapes mark the three regimes. Panel annotations give the correlation \emph{pooled} over the three regimes ($r=-0.57$ and $-0.59$ for degradation, $+0.02$ and $+0.01$ for the score); the regime-specific correlations quoted in \S\ref{sec:v2} ($-0.67$ to $-0.70$) are computed within each regime and are a different statistic. Predictor and outcome share no algebraic term (disjoint past/future windows); clustered CIs and the hierarchical model are in \S\ref{sec:v2}.}
\label{fig:9}
\end{figure*}

\textbf{Mechanism: update value tracks incumbent degradation, not detector score.} At each of 250 triggers the harness measured the incumbent's degradation over the five \emph{preceding} windows and the candidate-minus-incumbent BA over the five \emph{following} windows --- predictor and outcome share no algebraic term. Pre-trigger degradation predicts the future value of committing in every regime (regime-specific r $= -$0.67 to $-$0.70, every clustered CI95 away from zero; pooled over the three regimes r $= -$0.57, Fig.~\ref{fig:9}), while the detector's score at the same triggers shows no consistent association (regime-specific r $= -$0.01 to $+$0.05, every CI95 straddling zero; pooled $+$0.02).

A mixed-effects model controlling for severity and stream time confirms it: degradation is the only predictor whose CI95 consistently excludes zero ($\beta_{\mathrm{deg}} = -1.02$ [$-$1.61, $-$0.43] per BA point; VIF $\le$ 2.4), and the analysis replicates on the quantum-detector logs; within triggered decisions the evaluated detectors' scores carry no \emph{consistent} incremental signal, not provably none. The association is descriptive: triggers are nested in seeds and regimes, past and future performance are autocorrelated, and a bounded accuracy series gives a degraded incumbent more headroom mechanically, so this is \emph{predictive} support for the degradation--headroom account and does not identify a causal mechanism (Fig.~\ref{fig:9}; Online Resource~1, \S S1.1, \S S2.10).

\textbf{Supporting robustness.} \emph{Observed-data:} with every oracle privilege and leakage path removed (candidate, probe and detector recalibration drawn from observed past traffic only, without replacement from a value-deduplicated pool), the \textbf{leakage-free observed-data} arm reproduces the harm-regime rescue ($+$4.05 and $+$4.83 above always-deploy on ToN-IoT full and mild drift); it is the only tier \textbf{free of simulator-oracle information} (Online Resource~1, \S S2.7). \emph{Classifier and generator controls:} at 512/class the harmful zero-drift \emph{sign} appears for random forest, logistic regression, MLP and full-feature SVC and across replacement generators; size-matching removes it for the robust learners, localizing the residual magnitude to the fragile SVC-RBF pipeline (Online Resource~1, \S S1.6, \S S2.12). \emph{Quantum detector:} the harm, the gate's rescue and the per-trigger mechanism all replicate under a quantum-kernel monitor at a ${\sim}114\times$ simulated cost, a detector-invariance check rather than a quantum-advantage claim (Online Resource~1, \S S1.2).

\textbf{VBC-SG: what validating formally costs.} The sequential family of \S\ref{sec:riskgates} controls, under the conditional weak null of Proposition~1, the per-proposal probability of falsely establishing probe-superiority; it does not bound future deployment harm. A pre-specified budget frontier prices it (Online Resource~1, \S S2.12): the lifetime-budgeted pooled sequence recovers $93\%$ of always-deploying's benefit at cap 512 (${\approx}578$ adjudicated labels per proposal, deferrals included) \emph{under the approximate pooled analysis}, while the fully stratified VBC-SG-Cohort-sim, which carries the formal deployment-long guarantee, reaches $81\%$. The deployment-long stratified guarantee is thus \textbf{non-vacuous within the evaluated} balanced-probe budget, from $b{=}256$ onward, at an abstention of 74--86\%.

Empirically the risk-controlled gates committed \emph{zero observed} harmful updates over a complete 520-commit accounting (506 evaluable) against 39\% for always-deploy; these commits cluster within shared seeds and are \textbf{not treated as 506 independent Bernoulli trials}, so, being \emph{not independent trials}, they support no population-rate bound and ``zero observed'' is not ``eliminates''. A \textbf{pooled sequential gate provides useful empirical risk control} at tens of adjudicated labels, while the \textbf{formally aligned stratified guarantee} requires substantially larger balanced-probe budgets. All of these numbers are evidence under the historical frozen-transformer policy; the frontier was not re-evaluated under the size-matched self-contained controls, and nothing here shows VBC-SG to be \emph{necessary} in that setting (Online Resource~1, \S S2.9, \S S4).

\textbf{Chronological replays: the external boundary.} These replays ask whether the controlled-condition harm and the value of validation reappear on real timelines. Thirteen registered replays on real, chronologically ordered streams were evaluated in total: the seven of the pre-enumerated final matrix reported in Table~\ref{tab:chronological_q1} (seeds 601--630) and six earlier registered replays reported in Online Resource~1, \S S2.6. Each processes 200 windows of 256 flows strided to span the capture: ordering, labels and the temporal availability of every probe are real; the sampling density is not.

Every CICIDS incumbent collapses on a later window (no-adaptation BA 49--72\%) as each stream presents an unseen attack sub-type, and drift-triggered retraining recovers +13.6 to +43.8 BA points; on these deep-benefit streams the point and strict gates retain 73--100\% of that recovery (48\% for strict on the shallow Thursday intra-day split), paying a real, metric-dependent premium in balanced accuracy on three streams, and a pre-specified interventional test rules out probe class composition as the cause (probe \emph{staleness} remains the hypothesis consistent with the data). Where the incumbent stays healthy, on the two UNSW-NB15 timelines (82.9\% and 84.2\% BA), updating helps moderately and the point and strict gates have higher point estimates than always-deploy (strict $+2.78$ and $+1.34$, descriptive, committing 2.5 and 3.6 times per stream against 13.0 and 17.3), while the healthy Wednesday intra-day split is an unresolved counterexample with the gates $0.5$--$0.7$ points \emph{below} always-deploy. These are descriptive comparisons, not inferential superiority claims: the pre-specified chronological family tests strict against no-adaptation, a registration that is structurally easy to satisfy wherever the incumbent collapses, so its confirmatory content is confined to the healthy-incumbent replays.

\textbf{Net harm remains unobserved}: on none of the thirteen chronological streams does always-deploy lose to never adapting, so the harmful-update finding stays a controlled-condition result. This is an absence of observed harm in thirteen replays, not an estimate of harm frequency and not evidence that harm cannot occur; the replays estimate no deployment harm frequency, and ToN-IoT ships no timestamps, so the controlled harm benchmark has no chronological analogue. The price of that conservatism is equally visible in Table~\ref{tab:chronological_q1}: on the five CICIDS replays VBC-SG-Cohort-sim commits at most once per stream and retains only 0--9\% of always-deploy's recovery, whereas on the two healthy UNSW timelines it retains 91\% and 129\% at 1{,}232 and 1{,}634 labels per stream. Formal lifetime risk control can become extremely conservative exactly where the incumbent has the most recovery headroom; this is the operational price of the guarantee.

\input{tables_springer/table_chronological_q1.tex}

\textbf{Operational feasibility.} Every count above is an \emph{adjudicated} label; an operator pays in \emph{inspected flows}, and finding an \emph{attack} label at operating prevalence costs about $1/\pi$ inspections. A pre-specified acquisition-yield simulation prices that search: alert-enriched inspection cuts it 5--8$\times$ at every prevalence while the commit rule sees only an independent uniform validation sample (32 adjudications), so enrichment improves discovery without touching decision validity; the arm is bounded (candidate batch and detector calibration remain balanced) and is not an end-to-end deployment cost. A separate prevalence sensitivity shows the gate holds down to $\pi \approx 0.05$ but its harm-regime protection dissolves at $\pi = 0.01$, where two-thirds of 32-flow probes contain no attack; there the binding constraint is a starved \emph{trigger}, not the gate (Online Resource~1, \S S2.3). The historical frozen-policy diagnostic further shows that simply rejecting probe ties (\emph{reject-ties}) recovers most of the residual zero-drift loss at no additional labels (Online Resource~1, \S S2.12).

\section{Discussion}

The study --- historical diagnostic, ownership replication, zero-drift and full-drift size controls, common-harness comparison, exact-feature-disjoint sensitivity, chronological matrix --- answers three questions.

\textbf{Q1: How do challenger construction and evidence condition promotion conclusions?} Promotion conclusions, and the apparent ranking of update policies, are conditional on challenger construction and evidence; these factors should therefore be controlled, reported and interpreted explicitly. The sharpest form of the result is the ordering dependence: with evidence-disadvantaged 512-per-class challengers every label-free alternative appeared to help at zero drift, whereas at parity those gains vanished, and the pre-specified ordering-change rule fired for ATC and the ensemble in the source-row block and for ATC, DoC and the ensemble under exact-feature-disjoint roles (\S\ref{sec:baselines}, \S\ref{sec:value_disjoint}). A policy ranking obtained with one candidate generator need not survive another; this is a conditionality result, not a claim that any one construction is mandatory. Frozen incumbent-owned preprocessing accounted for the mean full-drift harm in its paired control (the ownership interaction reaches $+6.0$ points where the harm was reported), and the challengers' 4$\times$ nominal per-class evidence disadvantage accounts for the residual zero-drift mean harm (all size effects Holm-significant; matched-size mean effects compatible with $\pm0.5$ points in 3/3). Self-contained preprocessing is the controlled alternative evaluated here, not a universal production requirement, and nominal row-count parity is not effective information parity.

The full-drift source-row-disjoint control first found a homogeneous $+0.82$ to $+1.66$-point effect. The final exact-feature-disjoint sensitivity retains a positive resolved effect in all three ($+0.53$, $+1.67$, $+0.38$), but the ToN-IoT effect is sub-material; the defensible conclusion is therefore benchmark-dependent rather than homogeneous. In security terms the effect is a lower false-positive rate at approximately unchanged attack recall (\S\ref{sec:value_disjoint}), obtained on a benchmark pipeline whose absolute false-positive rates under full drift are not deployment-grade. The same block also yielded smaller absolute always-deploy gains on PortScan and ToN-IoT than the earlier independent block; being a between-block comparison, that change is descriptive. The controls identify representation and nominal row-count effects only within the evaluated pool construction; they do not establish effective information parity or attribute the between-design change causally to duplicates.

\textbf{Q2: When does validation add value?} Validation has its greatest observed average value when candidate construction or evidence is asymmetric or candidate quality is uncertain. Under the frozen transformer the gate partly compensated a representation bias; at self-contained 512/class it compensated a nominal-evidence disadvantage (up to $+1.68$ points, guardrail-clean in all but one cell). At matched 2{,}000-per-class evidence the source-row blocks show no positive resolved gate effect and one strict-gate cost; the exact-feature-disjoint B2 and B1 repetitions preserve that qualitative conclusion. The experiments therefore do not support universal validation, and we do not call validation necessary, generally safer, superior, or the recommended default.

Where it earns its keep, the gate works because it estimates a contemporaneous surrogate for the sign of the candidate--incumbent difference from a small labeled probe whose validity does not depend on the model being judged; the realized future value remains observable only downstream. A defensible \emph{operational} interpretation, which extrapolates beyond the fixed-policy evidence, is to validate when upstream comparability cannot be established cheaply or when tail-risk or governance requirements justify the caution, and to treat a demonstrably comparable, well-evidenced challenger as one that validation will mostly delay; we did not evaluate a meta-controller that selects a gate or its strength online.

The label bill is layered: retraining labels are unavoidable for any labeled-retraining loop; when candidate and validation labels are acquired jointly, reserving a pre-specified disjoint holdout of that same (deduplicated) labeled batch as the probe need not require label acquisition beyond the batch, at the opportunity cost of the held-out flows, which are not used to train the challenger; and converting average harm into average benefit in the controlled harm regime costs a fresh probe of $\approx$32 flows per confirmed drift under the controlled balanced-probe protocol, an adjudication count compatible with analyst-assisted workflows, though the inspected-flow acquisition cost of the balanced candidate batch and of the formal stratified probes was not evaluated. On collapsed chronological incumbents every conservative layer forfeits part of a large recovery, VBC-SG almost all of it (\S\ref{sec:boundaries}): where degradation is strong and well established, excessive validation delays beneficial adaptation, and the operator's protection is bounded deferral, adequate probe budgets and rollback capability rather than more gating.

\textbf{Q3: What does the common-harness comparison say about alternative update policies?} Different alternatives occupy different information/accuracy/update-cost trade-offs, and none dominates. Under exact-feature-disjoint roles ATC is unresolved on PortScan and UNSW and compatible on ToN-IoT at full drift, the calibrated ensemble is unresolved/material cost/compatible, DoC retains material costs on PortScan and UNSW, replay and DDM on UNSW, and ADWIN in all three (at the registered reference parameters for DDM/ADWIN), while ATC remains compatible with the point gate in five of six scenarios. These are statements about fixed evaluated configurations, not general properties of the methods.

\textbf{Implications for operational NIDS, and what is not implied.} Detection is not decision: a detector reports that the distribution moved, while promotion concerns the sign of the candidate--incumbent difference, with which the evaluated detector scores showed no consistent association within triggered decisions (regime-specific $r$ between $-0.01$ and $+0.05$ at 250 logged triggers, pooled $+0.02$; \S\ref{sec:v2}); the more expensive quantum-kernel monitor did not change promotion outcomes in its frozen-policy check. The controlled harm results license no prevalence claim about production, and the pool-constructed trajectories are not real deployment timelines. The contribution is evaluation discipline: challenger representation, training evidence and exact-value role separation should be stated before attributing harm, benefit or policy ordering, confined to the network-security benchmarks studied.

\textbf{Trust assumptions, security significance and scope.} The asset is the integrity and performance of the deployed intrusion detector; the decision is whether a challenger replaces the incumbent. In a deployed NIDS that promotion is a security-relevant integrity decision: it replaces the model responsible for all subsequent attack detection, so a poorly supported update can alter the detector's future security behaviour --- its attack recall and false-positive rate, which is why both are carried as guardrails throughout --- even when the drift alarm itself was correct. In every confirmatory block the candidate-training data, the trigger process and the validation evidence are assumed not to be adaptively manipulated by an adversary. Automated retraining does expose, in principle, an update surface to data manipulation: an adversary able to influence the traffic that raises alarms or trains challengers could steer what gets promoted, a risk studied in the poisoning literature \cite{biggio2012poisoning}. Adversarial candidate poisoning, trigger manipulation and adaptive attacks on the validation probe are out of scope: the present experiments do not model an adversary manipulating the update process --- candidate training data, drift triggers and validation probes are all assumed honest --- and the evaluated validation policies are not claimed as poisoning defenses. The only related evidence is a bounded, non-adaptive robustness probe from the initial study, in which uniformly random flips of up to 40\% of the probe labels left the gate harm-avoiding (Online Resource~1, \S S1.5); it does not model an adaptive adversary. The security contribution of this study is an evaluation discipline for a decision whose outcome determines the detector that will face the next attack.

\textbf{Operational guidance, not a validated stack.} The full combination --- a self-contained challenger, nominal size matching, observed-data construction, a detector alarm process, chronological deployment and natural prevalence --- was not evaluated jointly. The sequence below is therefore evaluation guidance and future-work hypotheses derived from separate controls, not a demonstrated or recommended deployment configuration:
\begin{enumerate}
  \item \emph{Audit candidate construction.} Do not treat incumbent-owned frozen preprocessing as a neutral evaluation convention; when comparability is intended, use a self-contained challenger, and if production deliberately freezes the incumbent's representation, state and validate that policy explicitly (\S\ref{sec:symmetric}).
  \item \emph{Audit evidence comparability.} Match the challenger's nominal training evidence to the incumbent's, or treat any shortfall as a reason to validate (\S\ref{sec:sizematched}, \S\ref{sec:sizematched_drift}); nominal parity is necessary for a fair reading, not sufficient for full comparability.
  \item \emph{Assess incumbent health.} Degradation, not detector score, predicts update value (\S\ref{sec:v2}); the cost and calibration of such monitoring were not evaluated here.
  \item \emph{Condition validation strength on uncertainty.} Validate conservatively when candidate quality or incumbent health is uncertain; when the challenger is comparable and degradation is strong and well established, avoid unnecessary delay, since conservative gates measurably forfeit recovery there (\S\ref{sec:boundaries}).
  \item \emph{Monitor post-promotion outcomes and retain rollback capability.} Mean compatibility within a margin is not a per-proposal guarantee (\S\ref{sec:v2}, \S\ref{sec:sizematched}).
\end{enumerate}
Under this design hypothesis, and conditional on a comparable proposal being generated, the detector acts mainly as a scheduler; at extreme class imbalance, detector calibration and alarm starvation determine whether any of this is invoked at all.

\textbf{Computational overhead of self-contained challengers (operational implication, not a benchmark).} A self-contained challenger refits its own preprocessing bundle (a feature standardizer and an 8-component PCA) on its candidate batch before training the classifier, and a commit deploys the complete bundle; inference cost after deployment is structurally similar under both policies. We did not run an isolated timing study of the two policies. The only measured figures are coarse per-arm wall-clock times of sequential single-machine runs: across the 18 matched scenario$\times$policy pairs of the symmetric-pipeline replication, self-contained arms took on average $0.97\times$ the wall-clock of their frozen counterparts (range $0.89$--$1.06$), so the transformer refit is not visible at whole-arm granularity, and the 2{,}000-per-class arms of the size-matched control took $1.2$--$1.5\times$ the wall-clock of their 512-per-class counterparts, consistent with SVC-RBF cost growing with the batch. These whole-arm timings do not isolate preprocessing overhead, and that a self-contained arm occasionally runs \emph{faster} than its frozen counterpart is run-to-run noise, not a speedup caused by self-contained preprocessing; they are operational context, not a benchmark, and distinct from the ${\sim}114\times$ simulated overhead of the quantum-kernel monitor (\S\ref{sec:v2}), which concerns the drift detector rather than the challenger pipeline.

\section{Limitations}

\textbf{Benchmarks and drift construction.} Three public benchmarks in binary benign-versus-attack form, with scenarios constructed from regime pools, do not constitute broad external validation; the strongest benefit evidence is concentrated in CICIDS2017, and the marginal and harmful ends come from UNSW-NB15 and ToN-IoT. The core drift trajectories are gradual mixing ramps between real regime pools (covariate/regime drift); we do not isolate a $p(y\,|\,x)$ change, and recurrent drift is untested. Candidate batches and probes are sampled at the true current severity, information a production system lacks; the leakage-free observed-data arm (\S\ref{sec:v2}; Online Resource~1, \S S2.7) reproduces the harmful-regime rescue from observed past traffic only, but equivalence to the oracle-probe arm was not tested and the size-matched controls were not repeated in observed-data form.

\textbf{Learners and pipelines.} The primary pipeline is SVC-RBF with an 8-dimensional PCA chosen for quantum-simulation tractability. At full drift the historical net-harm result is specific to that fragile classifier, the fragile-model tail of the evaluated learner spectrum, and does not appear for more robust classifiers (Online Resource~1, \S S1.6; \S\ref{sec:v2}); incremental or weighted online learners with no discrete replacement step remain untested, and the full-drift size control and the common-harness comparison were run on SVC-RBF only, so effect magnitudes are pipeline-dependent and generalization to other learners requires further evaluation. The pipeline's absolute false-positive rates under full drift (roughly 27--31\% on UNSW-Recon and ToN-IoT) are not deployment-grade; the study evaluates promotion behaviour under this benchmark pipeline, not a production-ready detector. Neither the frozen-policy full-drift harm nor the 512-challenger zero-drift harm should be quoted as a general property of drift-triggered updating: harm is what happens when a healthy model is replaced by a challenger whose construction or evidence is materially asymmetric, a condition on the proposal, not on the dataset.

\textbf{Nominal evidence is not effective information.} The size-matched controls equate \emph{nominal} per-class sample size; temporal coverage, diversity, subtype support, label quality, duplication, prevalence and effective sample size are not equated by equal row counts, and with-replacement candidate draws mean nominal rows can repeat even in the exact-feature-disjoint role design (\S\ref{sec:sizematched}). The final sensitivity prevents the same exact cleaned feature vector from crossing window/train/probe roles but preserves within-role multiplicity; it addresses exact duplicate-value exposure only, not approximate similarity, semantic duplicates or dependence among distinct vectors. The full-drift sensitivity establishes positive resolved size effects only within its nested pool design, with a material benefit in two of three benchmarks; we make no claim that larger candidates always help or that the effect is monotone beyond the two sizes. The zero-drift trigger is a random proposal rate, not a detector alarm process.

The controls isolate the design factors individually; the complete self-contained $+$ size-matched $+$ observed-data $+$ real-alarm $+$ natural-prevalence $+$ exact-feature-disjoint combination was not evaluated jointly, so their joint deployment behaviour is untested and the operational sequence in \S6 is a design hypothesis derived from separate controls. The contribution is evaluation methodology, not the validation of a deployment recipe. Mean compatibility at the matched size does not demonstrate absence of an effect, nor imply that no individual committed candidate can be worse than the incumbent; the future-negative sign percentages of \S\ref{sec:sizematched} are descriptive within-trajectory fractions over seed-clustered commits, not population prevalences or deployment probabilities.

\textbf{The common-harness comparison.} Information budgets differ legitimately by method definition (ATC and DoC use a 512-row training-time validation sample per model and zero target labels, DDM/ADWIN consume 800 monitoring labels per stream, the gates 32 target labels per decision) and are reported rather than equalized; policies with probability outputs train Platt-calibrated SVCs, which leaves the classifiers' predictions unchanged. DDM and ADWIN were run at their registered reference parameters and were not tuned; their cells characterize that configuration, and a parameter sweep would be required before reading them as properties of the methods. Recent adaptive-NIDS systems were reviewed as of September 2026 (Online Resource~1, \S S11); their triggers, labels, update semantics and decision units are not decision-equivalent to a per-proposal incumbent--challenger comparison, so they are reported as related systems rather than reproduced unfaithfully. Comparisons are between fixed policies; no learned meta-controller was evaluated. Cross-size contrasts are seed-paired only.

\textbf{Scope of the formal, diagnostic and chronological instruments.} The quantum detector, the full sequential frontier and the mild-drift matrix remain evaluated under the historical frozen-transformer policy; VBC-SG was not re-run under size-matched own pipelines, so none of those numbers should be assumed to transfer. VBC-SG's guarantee --- where it applies --- controls false probe-superiority \emph{on the probe distribution} under the conditional weak null of Proposition~1, not future deployment harm; the probe's fixed class and mixture quotas are not proved to convert a marginal non-superiority null into that conditional null, and the bound equals a bound on future harm only when the probe is representative, an assumption our own stale-probe and coverage results show can fail (the empirical-Bernstein sequence exceeds nominal $\alpha$ under strong within-probe autocorrelation), and its deployment-long budget is prohibitive in its maximally-stacked stratified form at $b{=}64$.

The gate assumes a small labeled probe at decision time; its size ($\approx$32--64 adjudicated flows per confirmed drift) is an incremental adjudication count under the evaluated protocol, not an estimate of inspected-flow acquisition cost in an operationally imbalanced deployment, which is not modeled; under genuinely random inspection the gate's harm-regime protection dissolves at $\pi = 0.01$, and the initial study's stronger claim, which rested on a probe conditioned to contain one attack, is not retained. Label latency is modeled for the probe (up to 20 windows stale) and the candidate batch (5--20 windows; Table~\ref{tab:latency}); asynchronous or partial labeling and candidate-aware label-flipping adversaries are untested. The thirteen chronological replays (seven in the registered final matrix, six earlier) are external boundary evidence: they provide no estimate or bound on how frequently harm occurs in deployment, most CICIDS streams collapse into deep-benefit territory, ToN-IoT ships no timestamps, so the controlled harm benchmark has no chronological counterpart, and the pre-specified chronological family is structurally easy to satisfy where the incumbent collapses. Quantum kernels are classically simulated, so the 114$\times$ monitoring overhead is a simulation figure.

\textbf{Statistical granularity and multiplicity.} The 30 seeds of each block estimate stochastic variability conditional on that benchmark and pool design; they are not 30 independent deployment environments. The external study units are the three benchmarks (seven regimes in the exploratory stage), and confidence intervals should be read at that granularity (Online Resource~1, \S S3 reports leave-one-out behaviour); the narrative therefore leads with effect sizes, intervals, heterogeneity across benchmarks and robustness across sensitivity blocks rather than with adjusted $p$-values. Multiplicity-adjusted $p$-values are computed directly from paired per-seed contrasts using a deterministic centered bootstrap, with Holm \cite{holm1979simple} within each pre-specified family of the core, construction, size, common-harness and exact-feature-disjoint blocks (Online Resource~1, \S S6--S11), and Benjamini--Hochberg \cite{benjamini1995controlling} separately for the pre-declared budget-frontier and chronological families. The exact-feature-disjoint B2-vs-historical contrasts use an independent-block bootstrap and Holm correction; they are sensitivity contrasts, not causal duplicate effects. Frozen-side cells, 512-vs-never columns and common-harness anchor contrasts remain descriptive and uncorrected.

\section{Conclusion}

A drift alarm can justify constructing a challenger, but not promoting it. In this study the apparent outcome of promotion, and the apparent ranking of update policies, changed with two upstream conditions. With incumbent-owned frozen preprocessing removed, the mean full-drift harm did not persist; with nominal candidate evidence matched to the incumbent's, the residual zero-drift deficit disappeared; and the pre-specified ordering-change rule fired for label-free and ensemble policies when the challenger's evidence changed. Under exact-cleaned-feature-disjoint roles and pool-constructed progressive drift, the 512$\to$2{,}000 intervention retained positive resolved effects of $+0.53$, $+1.67$ and $+0.38$ balanced-accuracy points, material in two of three benchmarks and driven mainly by fewer false positives.

At nominal parity, point and strict validation provided no resolved positive average effect, and strict validation incurred a sub-material cost in one cell. The common-harness interpretation is partially robust: policy ordering remains candidate-size-dependent and no evaluated policy globally dominates, while the earlier ATC and calibrated-ensemble compatibility claims narrow. Promotion conclusions are conditional on challenger construction and evidence; these factors should be controlled, reported and interpreted explicitly. Validation is conditional rather than universal, the net-harm findings are configuration-dependent, and thirteen chronological replays showed no net harm from always deploying while bounding neither its frequency nor its impossibility. The integrated operational stack remains future work rather than an empirically validated recommendation.

\section*{Statements and Declarations}
\paragraph{Funding} This research did not receive any specific grant from funding agencies in the public, commercial, or
not-for-profit sectors.

\paragraph{Competing interests} The authors declare that they have no known competing financial interests or personal relationships that
could have appeared to influence the work reported in this paper.

\paragraph{Ethics approval and consent to participate} Not applicable: the study uses three public benchmark datasets and involves no human participants or animals.

\paragraph{Consent for publication} Not applicable: the manuscript contains no data or images of individual persons.

\paragraph{Data availability} The three public benchmarks (CICIDS2017, UNSW-NB15, ToN-IoT) are cited in the text. Source code, frozen protocols, configs, environment specification, reproducibility instructions, confirmatory result summaries and the hash manifest are archived at Zenodo in artifact version v1.24.0: the exact version DOI is \href{https://doi.org/10.5281/zenodo.22239106}{10.5281/zenodo.22239106}, and the concept DOI \href{https://doi.org/10.5281/zenodo.21322256}{10.5281/zenodo.21322256} resolves to the latest version~\cite{fernandezbarrios2026artifact}. The previous v1.23.0 release remains immutable. Public benchmark datasets and voluminous raw/generated outputs are not redistributed; the released pipeline regenerates them, and every released result CSV used by the manuscript is pinned by SHA-256 in the manifest.

\paragraph{Code availability} The source code, frozen protocols, experiment runner and analysis scripts are part of the Zenodo artifact identified under Data availability.

\paragraph{Author contributions} Roberto Fern\'andez-Barrios: Conceptualization, Methodology, Software, Data curation, Formal analysis, Visualization, Investigation, Writing -- original draft, Writing -- review \& editing. Iker Pastor-L\'opez: Supervision, Validation, Writing -- review \& editing. Amaia Pikatza-Huerga: Validation, Project administration, Writing -- review \& editing. Pablo Garc\'ia Bringas: Supervision, Resources, Writing -- review \& editing.

\paragraph{Generative AI and AI-assisted technologies} During preparation of this work the authors used Claude (Anthropic) and OpenAI Codex to assist with drafting and editing text and with implementation and review of analysis scripts. The authors defined the protocols and scientific decisions, reviewed and verified all AI-assisted output, and take full responsibility for the work.

\bibliographystyle{spmpsci}  
\bibliography{references}
\end{document}

%% file: tables_springer/table_synthesis.tex
\begin{table*}[t]
\centering
\caption{\textbf{Central evidence matrix: how the apparent value of promotion depends on
challenger construction and evidence.} Each cell reports always-deploy minus never-adapt in
balanced-accuracy points (PortScan / UNSW-Recon / ToN-IoT) and whether point/strict validation
adds Holm-significant average value over always-deploy in that configuration; every value
names the 30-seed block it was estimated in, and where a cell's validation summary comes from
a different registered block than its $\Delta$, that block is named in the cell. Rows 1--2 are
the historical frozen incumbent-owned preprocessing configuration (descriptive; the frozen arm
of the symmetric replication, seeds 3001--3030, and, for the 2{,}000/class zero-drift cell,
the registered frozen size-matched control on seeds 104--133). Rows 3--4 are self-contained
challengers: zero-drift $\Delta$s from the registered size-matched control (seeds 4001--4030;
512 column descriptive, 2{,}000 column registered F1), full-drift $\Delta$s from the
registered size-matched control under drift (seeds 6001--6030; registered G1). Absolute
frozen-configuration magnitudes differ between seed blocks (\S\ref{sec:historical}); only
within-block contrasts are paired. ``not evaluated'' cells were never run under a registered
protocol.}
\label{tab:synthesis}
\footnotesize
\setlength{\tabcolsep}{4pt}
\renewcommand{\arraystretch}{1.15}
\begin{tabular}{p{0.20\linewidth} p{0.36\linewidth} p{0.36\linewidth}}
\toprule
Configuration & 512/class challenger & 2{,}000/class challenger \\
\midrule
Frozen incumbent-owned preprocessing (historical), full drift &
$\Delta$ vs never: $+6.31$ / $+1.39$ / $-4.95$ (seeds 3001--3030; descriptive)\newline validation: point $-$ naive $+0.33$ / $-0.00$ / $+5.56$ (same block; descriptive) &
not evaluated under a registered protocol \\
\addlinespace
Frozen incumbent-owned preprocessing (historical), zero drift &
$\Delta$ vs never: $-4.09$ / $-0.68$ / $-5.87$ (seeds 3001--3030; descriptive)\newline validation: point $-$ naive $+2.93$ / $+0.21$ / $+5.47$ (same block; descriptive) &
$\Delta$ vs never: $-4.81$ / $-0.13$ / $-5.76$ (seeds 104--133; descriptive)\newline validation: point $-$ naive $+4.46$ / $+0.03$ / $+5.35$ (same block; descriptive) \\
\addlinespace
Self-contained challenger, zero drift &
$\Delta$ vs never: $-1.70$ / $-0.65$ / $-0.24$ (seeds 4001--4030; descriptive)\newline validation: 6/6 gate contrasts Holm-significant gains (seeds 3001--3030, Table~\ref{tab:symmetric_pipeline}) &
$\Delta$ vs never: $+0.19$ / $-0.02$ / $-0.01$ (seeds 4001--4030; registered F1; CI90 within $\pm0.5$ in 3/3)\newline validation: 0/6 Holm-significant gains (same block; registered F3) \\
\addlinespace
Self-contained challenger, full drift &
$\Delta$ vs never: $+8.90$ / $+2.13$ / $+1.44$ (seeds 6001--6030; registered G1)\newline validation: 1/6 Holm-significant gain, 1 resolved cost (seeds 3001--3030, Table~\ref{tab:symmetric_pipeline}) &
$\Delta$ vs never: $+9.72$ / $+3.79$ / $+2.44$ (seeds 6001--6030; registered G1)\newline validation: 0/6 Holm-significant gains, 1 resolved cost (strict, UNSW $-0.34$; same block; registered G3) \\
\bottomrule
\end{tabular}
\end{table*}

%% file: tables_springer/table_symmetric_pipeline.tex
\begin{table*}
\centering
\caption{\textbf{The symmetric-pipeline dynamic replication (registered; seeds 3001--3030).}
Shared raw stream per seed (hash-verified across all seven arms); \emph{frozen} = historical
frozen-initial-transformer policy, \emph{own} = self-contained pipelines (each challenger
fits its own scaler/PCA on its own raw batch, bit-identical across policies; commits deploy
the complete bundle). BA points, paired within seed, 30 seeds, deterministic paired-bootstrap
CI95; $\dagger$ = Holm-significant within its registered family (own rows); frozen rows are
descriptive (uncorrected). Zero drift: random trigger $p{=}0.05$, no drift; the detector
representation stays the initial transformer under both policies.}
\label{tab:symmetric_pipeline}
\small
\resizebox{\ifdim\width>\linewidth \linewidth\else\width\fi}{!}{%
\begin{tabular}{l l r l l l}
\toprule
Regime & Transf. & Never BA & Naive $-$ never & Point $-$ naive & Strict $-$ naive \\
\midrule
PortScan full & frozen & 87.44 & $+6.31$ [4.11, 8.53] & $+0.33$ [-0.05, 1.00] & $+0.36$ [-0.12, 1.07] \\
 & own & 87.44 & $+7.21$ [4.89, 9.52]$^{\dagger}$ & $+0.19$ [-0.06, 0.41] & $+0.05$ [-0.33, 0.41] \\
\addlinespace
UNSW-Recon full & frozen & 83.03 & $+1.39$ [1.17, 1.62] & $-0.00$ [-0.11, 0.10] & $-0.03$ [-0.24, 0.16] \\
 & own & 83.03 & $+2.55$ [2.34, 2.75]$^{\dagger}$ & $-0.21$ [-0.36, -0.06]$^{\dagger}$ & $-0.15$ [-0.35, 0.05] \\
\addlinespace
ToN-IoT full & frozen & 92.40 & $-4.95$ [-7.15, -2.86] & $+5.56$ [3.50, 7.75] & $+5.38$ [3.34, 7.56] \\
 & own & 92.40 & $+1.03$ [0.55, 1.53]$^{\dagger}$ & $+0.64$ [0.26, 1.06]$^{\dagger}$ & $+0.18$ [-0.31, 0.67] \\
\addlinespace
PortScan zero & frozen & 95.44 & $-4.09$ [-6.01, -2.56] & $+2.93$ [1.34, 4.84] & $+4.03$ [2.49, 5.95] \\
 & own & 95.44 & $-1.74$ [-2.35, -0.96]$^{\dagger}$ & $+0.75$ [0.41, 1.16]$^{\dagger}$ & $+1.68$ [1.23, 2.13]$^{\dagger}$ \\
\addlinespace
UNSW-Recon zero & frozen & 89.09 & $-0.68$ [-0.80, -0.57] & $+0.21$ [0.12, 0.32] & $+0.44$ [0.31, 0.57] \\
 & own & 89.09 & $-0.65$ [-0.77, -0.55]$^{\dagger}$ & $+0.19$ [0.09, 0.31]$^{\dagger}$ & $+0.51$ [0.38, 0.64]$^{\dagger}$ \\
\addlinespace
ToN-IoT zero & frozen & 92.58 & $-5.87$ [-9.20, -3.29] & $+5.47$ [2.79, 8.86] & $+5.87$ [3.29, 9.21] \\
 & own & 92.58 & $-0.38$ [-0.53, -0.25]$^{\dagger}$ & $+0.11$ [0.03, 0.20]$^{\dagger}$ & $+0.34$ [0.20, 0.49]$^{\dagger}$ \\
\bottomrule
\end{tabular}}
\end{table*}

%% file: tables_springer/table_size_matched.tex
\begin{table*}
\centering
\caption{\textbf{The size-matched self-contained challenger control (registered; seeds
4001--4030).} Zero drift, random proposal trigger, own-transformer only; the sole new
variable is the challenger's per-class training size (512 vs.\ the incumbent's 2{,}000),
with nested candidate batches (\S\ref{sec:sizematched_method}). BA points, paired within
seed, 30 fresh seeds, paired-bootstrap CI95; $\dagger$ = Holm-significant within its
registered family; the 512-vs-never column is descriptive (uncorrected). \emph{compat.}:
CI90 fully inside the preregistered $\pm0.5$-point margin; PortScan is boundary-close
(upper bound $0.494$, $0.006$ inside) and does not demonstrate absence of an effect.}
\label{tab:size_matched}
\small
\resizebox{\ifdim\width>\linewidth \linewidth\else\width\fi}{!}{%
\begin{tabular}{l r l l c l l l}
\toprule
Regime & Never BA & Naive$_{512}-$never & Naive$_{2000}-$never & compat. &
Naive$_{2000}-$naive$_{512}$ & Point$_{2000}-$naive$_{2000}$ &
Strict$_{2000}-$naive$_{2000}$ \\
\midrule
PortScan zero & 95.66 & $-1.70$ [-2.05, -1.36] & $+0.19$ [-0.15, 0.55] & yes & $+1.89$ [1.54, 2.23]$^{\dagger}$ & $+0.05$ [-0.00, 0.12] & $+0.13$ [-0.07, 0.34] \\
UNSW-Recon zero & 89.08 & $-0.65$ [-0.77, -0.55] & $-0.02$ [-0.10, 0.06] & yes & $+0.63$ [0.53, 0.73]$^{\dagger}$ & $+0.01$ [-0.02, 0.04] & $+0.06$ [-0.01, 0.12] \\
ToN-IoT zero & 92.56 & $-0.24$ [-0.39, -0.10] & $-0.01$ [-0.05, 0.04] & yes & $+0.23$ [0.10, 0.38]$^{\dagger}$ & $+0.02$ [0.00, 0.04] & $+0.03$ [-0.01, 0.08] \\
\bottomrule
\end{tabular}}
\end{table*}

%% file: tables_springer/table_size_matched_drift.tex
\begin{table*}[t]
\centering
\caption{\textbf{Candidate evidence under pool-constructed progressive drift: the registered
source-row-disjoint size-matched control (seeds 6001--6030, 21 arms).} Self-contained challengers, nested
candidate batches drawn at the proposal-time mixture (the 512 batch is the first 512 rows
per class of the 2{,}000 batch). BA points; paired within seed, 30 seeds; CI95 from the
deterministic centered paired bootstrap; $\dagger$ = Holm-significant within its registered
family (G1: naive vs never at both sizes; G2: the primary size effect; G3: gate value at
2{,}000). Historical registered classification of G2 per the frozen protocol; block outcome
\textsc{homogeneous--size benefit}. Columns 3--4: always-deploy
minus never-adapt at each size; column 5: the size effect; columns 7--8: gate minus
always-deploy at 2{,}000/class. Full matrices in Online Resource~1, \S S9.}
\label{tab:size_matched_drift}
\footnotesize
\setlength{\tabcolsep}{3pt}
\begin{tabular}{l r r r l l l l}
\toprule
Regime (full drift) & Never BA & Naive$_{512}$ & Naive$_{2000}$ &
Size effect [CI95] & G2 class & Point$_{2000}$ [CI95] & Strict$_{2000}$ [CI95] \\
\midrule
PortScan & 85.58 & $+8.90$$^{\dagger}$ & $+9.72$$^{\dagger}$ & $+0.82$ [0.67, 0.98]$^{\dagger}$ & size benefit & $+0.15$ [-0.03, 0.39] & $-0.00$ [-0.17, 0.23] \\
UNSW-Recon & 83.36 & $+2.13$$^{\dagger}$ & $+3.79$$^{\dagger}$ & $+1.66$ [1.51, 1.81]$^{\dagger}$ & size benefit & $-0.06$ [-0.16, 0.03] & $-0.34$ [-0.53, -0.15]$^{\dagger}$ \\
ToN-IoT & 91.79 & $+1.44$$^{\dagger}$ & $+2.44$$^{\dagger}$ & $+1.00$ [0.60, 1.39]$^{\dagger}$ & size benefit & $+0.01$ [-0.06, 0.09] & $-0.14$ [-0.56, 0.28] \\
\bottomrule
\end{tabular}
\end{table*}

%% file: tables_springer/table_evidence_validation_tradeoff.tex
\begin{table*}[t]
\centering
\caption{Nominal evidence--validation trade-off under the preregistered zero-drift size-matched control (own-transformer, random proposals, balanced pools, SVC-RBF, seeds 4001--4030). ``Cand.''/``Probe'' are nominal adjudicated labels per proposal; moving from 512 to 2{,}000 per class adds $2\times1{,}488 = 2{,}976$ candidate-training labels per proposal versus the 32-label point/strict probe --- counts, not costs: inspected-flow acquisition is not modeled and no economic dominance is claimed. Paired per-seed BA contrasts (CI95, deterministic centered paired bootstrap): naive-2{,}000 vs never is registered F1, gate gains at 2{,}000 are registered F3 (none Holm-significant), 512-side and gate-vs-never cells are descriptive, uncorrected. The supplementary point/strict-2{,}000 policies differ from naive-2{,}000 by $<0.14$ pp (non-significant, CI90 compatible within $\pm$0.5 pp). Guardrails: all matched-size gate cells pass recall/FPR non-inferiority; at 512 the UNSW strict cell fails recall non-inferiority (Online Resource~1, \S S8).}
\label{tab:evidence_validation_tradeoff}
\footnotesize
\setlength{\tabcolsep}{3.5pt}
\resizebox{\ifdim\width>\linewidth \linewidth\else\width\fi}{!}{%
\begin{tabular}{l l r r l l r}
\toprule
Dataset & Policy & Cand. & Probe & BA vs never [CI95] & Gate gain vs naive [CI95] & Commits/seed \\
\midrule
\multirow{4}{*}{CICIDS2017-PortScan} & naive 512 & 1,024 & 0 & $-1.70$ [-2.05, -1.36] & --- & 3.70 \\
 & point 512 + 32-label probe & 1,024 & 32 & $-0.73$ [-1.10, -0.38] & $+0.97$ [+0.67, +1.31] & 1.87 \\
 & strict 512 + 32-label probe & 1,024 & 32 & $-0.03$ [-0.24, +0.16] & $+1.67$ [+1.36, +2.00] & 0.23 \\
 & naive 2{,}000 & 4,000 & 0 & $+0.19$ [-0.15, +0.55] & --- & 3.70 \\
\midrule
\multirow{4}{*}{UNSW-NB15} & naive 512 & 1,024 & 0 & $-0.65$ [-0.77, -0.55] & --- & 3.70 \\
 & point 512 + 32-label probe & 1,024 & 32 & $-0.47$ [-0.61, -0.33] & $+0.19$ [+0.10, +0.30] & 2.27 \\
 & strict 512 + 32-label probe & 1,024 & 32 & $-0.19$ [-0.30, -0.10] & $+0.46$ [+0.35, +0.58] & 0.67 \\
 & naive 2{,}000 & 4,000 & 0 & $-0.02$ [-0.10, +0.06] & --- & 3.70 \\
\midrule
\multirow{4}{*}{ToN-IoT} & naive 512 & 1,024 & 0 & $-0.24$ [-0.39, -0.10] & --- & 3.70 \\
 & point 512 + 32-label probe & 1,024 & 32 & $-0.12$ [-0.24, -0.01] & $+0.12$ [+0.04, +0.23] & 3.13 \\
 & strict 512 + 32-label probe & 1,024 & 32 & $+0.00$ [-0.08, +0.06] & $+0.24$ [+0.12, +0.38] & 0.27 \\
 & naive 2{,}000 & 4,000 & 0 & $-0.01$ [-0.05, +0.04] & --- & 3.70 \\
\bottomrule
\end{tabular}}
\end{table*}

%% file: tables_springer/table_common_harness.tex
\begin{table*}[t]
\centering
\caption{\textbf{Registered common-harness comparison with published and reference baselines
(primary condition: self-contained 2{,}000/class challengers; seeds 5001--5030).} Balanced-accuracy
points over never-adapting (from the sealed per-arm means; the never-adapt row gives its absolute
mean BA; all arms share bit-identical raw streams per seed). Markers record the registered per-cell classification of the policy$-$naive
contrast (families PF1 zero drift / PF2 full drift; Holm within family): $\dagger$ material
gain, $\ddagger$ material cost, $\approx$ compatible (CI90 within $\pm0.5$), $?$ unresolved.
Anchor rows (naive, point, strict) are descriptive here by amendment. Labels = target labels
per decision; ATC/DoC additionally use a 512-row labeled validation sample at each model's
training time; DDM/ADWIN consume 8 monitoring labels per window. Origin: published generic
method, reference implementation (\texttt{river} 0.25.0), standard baseline, or authors'
policy --- none is an adaptive-NIDS system reproduced end to end. Full families, the 512/class
sensitivity block and statements in Online Resource~1, \S S10.}
\label{tab:common_harness}
\scriptsize
\setlength{\tabcolsep}{2pt}
\renewcommand{\arraystretch}{1.08}
\begin{tabular}{l l l r r r r r r}
\toprule
 & & & \multicolumn{3}{c}{Full drift} & \multicolumn{3}{c}{Zero drift} \\
\cmidrule(lr){4-6}\cmidrule(lr){7-9}
Policy & Origin & Labels at decision & PortScan & UNSW & ToN-IoT & PortScan & UNSW & ToN-IoT \\
\midrule
Never-adapt & anchor & 0 & 86.0 & 83.1 & 91.9 & 95.1 & 89.0 & 92.5 \\
Always-deploy (naive) & anchor & 0 & $+9.44$ & $+4.06$ & $+2.50$ & $+0.25$ & $+0.07$ & $+0.02$ \\
Point gate, $b{=}32$ & authors' policy & 32 & $+9.53$ & $+4.05$ & $+2.70$ & $+0.49$ & $+0.09$ & $+0.03$ \\
Strict gate (reject ties), $b{=}32$ & authors' policy & 32 & $+9.28$ & $+3.83$ & $+2.28$ & $+0.64$ & $+0.06$ & $-0.01$ \\
ATC \cite{garg2022atc} & published generic & 0 ($+$512-row val.) & $+8.80$$^{?}$ & $+3.84$$^{\approx}$ & $+2.60$$^{\approx}$ & $+0.52$$^{\approx}$ & $+0.03$$^{\approx}$ & $+0.02$$^{\approx}$ \\
DoC \cite{guillory2021doc} & published generic & 0 ($+$512-row val.) & $+6.85$$^{\ddagger}$ & $+3.47$$^{\ddagger}$ & $+2.71$$^{\approx}$ & $+0.64$$^{?}$ & $+0.06$$^{\approx}$ & $+0.05$$^{\approx}$ \\
Calibrated soft ensemble & standard baseline & 0 & $+9.35$$^{\approx}$ & $+3.64$$^{\approx}$ & $+2.26$$^{\approx}$ & $+1.71$$^{\dagger}$ & $+0.12$$^{\approx}$ & $+0.04$$^{\approx}$ \\
Replay 50/50 retraining & standard baseline & 0 & $+8.55$$^{\ddagger}$ & $+1.99$$^{\ddagger}$ & $+1.28$$^{\ddagger}$ & $+0.40$$^{\approx}$ & $+0.05$$^{\approx}$ & $+0.01$$^{\approx}$ \\
river-DDM trigger & reference impl. & 0; 800/stream monitoring & $+8.66$$^{\ddagger}$ & $+1.39$$^{\ddagger}$ & $+0.69$$^{\ddagger}$ & $+0.40$$^{?}$ & $+0.00$$^{\approx}$ & $-0.06$$^{\approx}$ \\
river-ADWIN trigger & reference impl. & 0; 800/stream monitoring & $+0.71$$^{\ddagger}$ & $+0.00$$^{\ddagger}$ & $+0.00$$^{\ddagger}$ & $+0.00$$^{?}$ & $+0.00$$^{\approx}$ & $+0.00$$^{\approx}$ \\
\bottomrule
\end{tabular}
\end{table*}

%% file: tables_springer/table_value_disjoint_main.tex
\begin{table*}[t]
\centering
\caption{\textbf{Exact-feature-disjoint sensitivity of the full-drift size effect.}
Always-deploy with 2{,}000/class minus always-deploy with 512/class, balanced-accuracy
points, 30 fresh seeds (7001--7030). Exact cleaned raw feature groups are assigned wholly
to window, candidate-training or probe roles; multiplicity and original labels are retained.
CI95 and Holm-adjusted $p$ are from the registered deterministic paired bootstrap. The
historical source-row-disjoint estimate is shown for context; ``change'' is sensitivity
minus historical, estimated between independent seed blocks and is not a causal duplicate-
leakage effect. The last two columns give always-deploy attack recall and false-positive
rate (\%) at 512 $\to$ 2{,}000 per class from the same sealed cells: the balanced-accuracy
advantage is driven mainly by the lower false-positive rate, with recall approximately
stable and slightly lower on UNSW-Recon.}
\label{tab:value_disjoint_main}
\scriptsize
\setlength{\tabcolsep}{2.5pt}
\begin{tabular}{l r l r l l l l l}
\toprule
Benchmark & Historical & Exact-feature-disjoint [CI95] & $p_{\mathrm{Holm}}$ & Class & Change [CI95] & Change class & Recall 512$\to$2{,}000 & FPR 512$\to$2{,}000 \\
\midrule
PortScan & $+0.82$ & $+0.53$ [0.27, 0.79] & 0.00018 & material benefit & $-0.29$ [-0.60, 0.01] & unresolved & 89.82$\to$90.13 & 5.44$\to$4.70 \\
UNSW-Recon & $+1.66$ & $+1.67$ [1.47, 1.89] & 0.00003 & material benefit & $+0.01$ [-0.24, 0.28] & compatible & 97.86$\to$97.51 & 31.20$\to$27.51 \\
ToN-IoT & $+1.00$ & $+0.38$ [0.21, 0.57] & 0.00018 & resolved sub-material & $-0.62$ [-1.04, -0.19] & material attenuation & 97.16$\to$97.34 & 28.39$\to$27.81 \\
\bottomrule
\end{tabular}
\end{table*}

%% file: tables_springer/table_chronological_q1.tex
\begin{table*}[t]
\centering
\caption{\textbf{The registered chronological matrix.} Seven pre-enumerated
time-ordered replays (CICIDS2017 train$\to$future day pairs, two intra-day splits, UNSW-NB15
at two training fractions), 200 windows of 256 flows in capture order, seeds 601--630.
Columns: balanced-accuracy points over never-adapting; \emph{no-adapt} is the frozen
incumbent's absolute BA (the health indicator). \textbf{No stream shows net harm.} Where
the incumbent collapses (CICIDS), the gates preserve part of always-deploying's recovery
but can impose substantial costs; where the incumbent stays healthy, point estimates for
point and strict gates are \emph{above} always-deploying on both UNSW
timelines (VBC-SG on the 20\% split but not the 40\%), with the healthy Wednesday intra-day
replay the unresolved counterexample. These are descriptive comparisons, not direct
inferential superiority claims.}
\label{tab:chronological_q1}
\small
\begin{tabular}{l r r r r r}
\toprule
Replay & no-adapt BA & naive & point & strict & VBC-SG \\
\midrule
Tue $\to$ Wed+Thu+Fri & 59.0 & $+14.78$ & $+11.52$ & $+11.50$ & $+0.16$ \\
Wed $\to$ Fri & 51.8 & $+35.29$ & $+27.64$ & $+28.81$ & $+0.91$ \\
Thu $\to$ Fri & 49.0 & $+36.07$ & $+27.71$ & $+26.51$ & $+1.56$ \\
Wed intra-day & 84.7 & $+6.47$ & $+6.02$ & $+5.77$ & $+0.56$ \\
Thu intra-day & 48.3 & $+0.88$ & $+0.88$ & $+0.42$ & $+0.00$ \\
UNSW (train 20\%) & 82.9 & $+8.35$ & $+9.90$ & $+11.13$ & $+10.76$ \\
UNSW (train 40\%) & 84.2 & $+7.64$ & $+8.38$ & $+8.98$ & $+6.92$ \\
\bottomrule
\end{tabular}
\end{table*}

%% file: main_springer.bbl
\begin{thebibliography}{10}
\providecommand{\url}[1]{{#1}}
\providecommand{\urlprefix}{URL }
\expandafter\ifx\csname urlstyle\endcsname\relax
  \providecommand{\doi}[1]{DOI~\discretionary{}{}{}#1}\else
  \providecommand{\doi}{DOI~\discretionary{}{}{}\begingroup
  \urlstyle{rm}\Url}\fi

\bibitem{wahab2022intrusion}
Abdel~Wahab, O.: Intrusion detection in the {IoT} under data and concept
  drifts: Online deep learning approach.
\newblock IEEE Internet of Things Journal \textbf{9}(20), 19706--19716 (2022).
\newblock \doi{10.1109/JIOT.2022.3167005}

\bibitem{aburomman2026automl}
Aburomman, A.A., Reaz, M.B.I.: {AutoML} for network-based intrusion detection:
  Evaluation practice, dataset quality, and deployment constraints.
\newblock Future Internet \textbf{18}(8), 383 (2026).
\newblock \doi{10.3390/fi18080383}

\bibitem{aguiar2024locality}
Aguiar, G.J., Cano, A.: A comprehensive analysis of concept drift locality in
  data streams.
\newblock Knowledge-Based Systems \textbf{289}, 111535 (2024).
\newblock \doi{10.1016/j.knosys.2024.111535}

\bibitem{alsaedi2020toniot}
Alsaedi, A., Moustafa, N., Tari, Z., Mahmood, A., Anwar, A.: {TON\_IoT}
  telemetry dataset: A new generation dataset of {IoT} and {IIoT} for
  data-driven intrusion detection systems.
\newblock IEEE Access \textbf{8}, 165130--165150 (2020).
\newblock \doi{10.1109/ACCESS.2020.3022862}

\bibitem{amalapuram2024spider}
Amalapuram, S.K., Tamma, B.R., Channappayya, S.S.: {SPIDER}: A semi-supervised
  continual learning-based network intrusion detection system.
\newblock In: IEEE INFOCOM 2024 -- IEEE Conference on Computer Communications,
  pp. 571--580. IEEE (2024).
\newblock \doi{10.1109/INFOCOM52122.2024.10621428}

\bibitem{andresini2021insomnia}
Andresini, G., Pendlebury, F., Pierazzi, F., Loglisci, C., Appice, A.,
  Cavallaro, L.: {INSOMNIA}: Towards concept-drift robustness in network
  intrusion detection.
\newblock In: Proceedings of the 14th ACM Workshop on Artificial Intelligence
  and Security (AISec), pp. 111--122 (2021).
\newblock \doi{10.1145/3474369.3486864}

\bibitem{apruzzese2023sok}
Apruzzese, G., Laskov, P., Schneider, J.: {SoK}: Pragmatic assessment of
  machine learning for network intrusion detection.
\newblock In: 2023 IEEE 8th European Symposium on Security and Privacy
  (EuroS\&P), pp. 592--614. IEEE (2023).
\newblock \doi{10.1109/EuroSP57164.2023.00042}

\bibitem{arp2022dodonts}
Arp, D., Quiring, E., Pendlebury, F., Warnecke, A., Pierazzi, F., Wressnegger,
  C., Cavallaro, L., Rieck, K.: Dos and don'ts of machine learning in computer
  security.
\newblock In: 31st USENIX Security Symposium (USENIX Security 22), pp.
  3971--3988 (2022)

\bibitem{baek2022agreement}
Baek, C., Jiang, Y., Raghunathan, A., Kolter, J.Z.: Agreement-on-the-line:
  Predicting the performance of neural networks under distribution shift.
\newblock In: Advances in Neural Information Processing Systems (NeurIPS), pp.
  19274--19289 (2022)

\bibitem{baena2006early}
Baena-Garc{\'i}a, M., del Campo-{\'A}vila, J., Fidalgo-Merino, R., Bifet, A.,
  Gavald{\`a}, R., Morales-Bueno, R.: Early drift detection method.
\newblock In: 4th International Workshop on Knowledge Discovery from Data
  Streams (IWKDDS) (2006)

\bibitem{bakirov2021automated}
Bakirov, R., Fay, D., Gabrys, B.: Automated adaptation strategies for stream
  learning.
\newblock Machine Learning \textbf{110}(6), 1429--1462 (2021).
\newblock \doi{10.1007/s10994-021-05992-x}

\bibitem{bansal2019updates}
Bansal, G., Nushi, B., Kamar, E., Weld, D.S., Lasecki, W.S., Horvitz, E.:
  Updates in human-ai teams: Understanding and addressing the
  performance/compatibility tradeoff.
\newblock In: Proceedings of the AAAI Conference on Artificial Intelligence,
  vol.~33, pp. 2429--2437 (2019).
\newblock \doi{10.1609/aaai.v33i01.33012429}

\bibitem{barbero2022transcendent}
Barbero, F., Pendlebury, F., Pierazzi, F., Cavallaro, L.: Transcending
  {TRANSCEND}: Revisiting malware classification in the presence of concept
  drift.
\newblock In: 2022 IEEE Symposium on Security and Privacy (S\&P), pp. 805--823
  (2022).
\newblock \doi{10.1109/SP46214.2022.9833659}

\bibitem{bayram2022degradation}
Bayram, F., Ahmed, B.S., Kassler, A.: From concept drift to model degradation:
  An overview on performance-aware drift detectors.
\newblock Knowledge-Based Systems \textbf{245}, 108632 (2022).
\newblock \doi{10.1016/j.knosys.2022.108632}

\bibitem{bellante2025evaluating}
Bellante, A., Fioravanti, T., Carminati, M., Zanero, S., Luongo, A.: Evaluating
  the potential of quantum machine learning in cybersecurity: A case-study on
  {PCA}-based intrusion detection systems.
\newblock Computers \& Security \textbf{154}, 104341 (2025).
\newblock \doi{10.1016/j.cose.2025.104341}

\bibitem{benjamini1995controlling}
Benjamini, Y., Hochberg, Y.: Controlling the false discovery rate: A practical
  and powerful approach to multiple testing.
\newblock Journal of the Royal Statistical Society: Series B (Methodological)
  \textbf{57}(1), 289--300 (1995).
\newblock \doi{10.1111/j.2517-6161.1995.tb02031.x}

\bibitem{bifet2007learning}
Bifet, A., Gavald{\`a}, R.: Learning from time-changing data with adaptive
  windowing.
\newblock In: Proceedings of the 2007 SIAM International Conference on Data
  Mining (SDM), pp. 443--448 (2007).
\newblock \doi{10.1137/1.9781611972771.42}

\bibitem{biggio2012poisoning}
Biggio, B., Nelson, B., Laskov, P.: Poisoning attacks against support vector
  machines.
\newblock In: Proceedings of the 29th International Conference on Machine
  Learning (ICML), pp. 1467--1474 (2012).
\newblock ArXiv:1206.6389

\bibitem{cacciarelli2024active}
Cacciarelli, D., Kulahci, M.: Active learning for data streams: a survey.
\newblock Machine Learning \textbf{113}, 185--239 (2024).
\newblock \doi{10.1007/s10994-023-06454-2}

\bibitem{camarda2025managing}
Camarda, F., De~Paola, A., Drago, S., Ferraro, P., Lo~Re, G.: Managing concept
  drift in online intrusion detection systems with active learning.
\newblock In: Proceedings of ITASEC \& SERICS 2025, CEUR Workshop Proceedings,
  vol. 3962 (2025)

\bibitem{chen2023continuous}
Chen, Y., Ding, Z., Wagner, D.: Continuous learning for {Android} malware
  detection.
\newblock In: 32nd USENIX Security Symposium (USENIX Security 23), pp.
  1127--1144. USENIX Association (2023).
\newblock
  \urlprefix\url{https://www.usenix.org/conference/usenixsecurity23/presentation/chen-yizheng}

\bibitem{constantinides2019novel}
Constantinides, C., Shiaeles, S., Ghita, B., Kolokotronis, N.: A novel online
  incremental learning intrusion prevention system.
\newblock In: 2019 10th IFIP International Conference on New Technologies,
  Mobility and Security (NTMS) (2019).
\newblock \doi{10.1109/NTMS.2019.8763842}

\bibitem{darling1957ks}
Darling, D.A.: The {Kolmogorov--Smirnov}, {Cram{\'e}r--von Mises} tests.
\newblock The Annals of Mathematical Statistics \textbf{28}(4), 823--838
  (1957).
\newblock \doi{10.1214/aoms/1177706788}

\bibitem{depaola2026hoids}
De~Paola, A., Drago, S., Ferraro, P., Lo~Re, G.: {HOIDS}: Concept drift aware
  hybrid online intrusion detection system.
\newblock Journal of Network and Computer Applications \textbf{254}, 104556
  (2026).
\newblock \doi{10.1016/j.jnca.2026.104556}

\bibitem{deng2021labels}
Deng, W., Zheng, L.: Are labels always necessary for classifier accuracy
  evaluation?
\newblock In: Proceedings of the IEEE/CVF Conference on Computer Vision and
  Pattern Recognition (CVPR), pp. 15069--15078 (2021)

\bibitem{engelen2021troubleshooting}
Engelen, G., Rimmer, V., Joosen, W.: Troubleshooting an intrusion detection
  dataset: the {CICIDS2017} case study.
\newblock In: 2021 IEEE Security and Privacy Workshops (SPW), pp. 7--12 (2021).
\newblock \doi{10.1109/SPW53761.2021.00009}

\bibitem{fernandezbarrios2026artifact}
Fern{\'a}ndez-Barrios, R., Pastor-L{\'o}pez, I., Pikatza-Huerga, A.,
  Garc{\'i}a~Bringas, P.: Validate before commit: reproducibility artifact
  (code and protocols).
\newblock Zenodo [software]; concept DOI 10.5281/zenodo.21322256 resolves to
  the latest version (2026).
\newblock \doi{10.5281/zenodo.22239106}.
\newblock
  \urlprefix\url{https://github.com/roberto-fernandez-barrios/Validate-Before-Commit/tree/v1.24.0}.
\newblock Version 1.24.0

\bibitem{flood2024smells}
Flood, R., Engelen, G., Aspinall, D., Desmet, L.: Bad design smells in
  benchmark {NIDS} datasets.
\newblock In: 2024 IEEE 9th European Symposium on Security and Privacy
  (EuroS\&P), pp. 658--675. IEEE (2024).
\newblock \doi{10.1109/EuroSP60621.2024.00042}

\bibitem{gama2004learning}
Gama, J., Medas, P., Castillo, G., Rodrigues, P.: Learning with drift
  detection.
\newblock In: Advances in Artificial Intelligence -- SBIA 2004, pp. 286--295
  (2004).
\newblock \doi{10.1007/978-3-540-28645-5_29}

\bibitem{gama2014survey}
Gama, J., {\v{Z}}liobait{\.e}, I., Bifet, A., Pechenizkiy, M., Bouchachia, A.:
  A survey on concept drift adaptation.
\newblock ACM Computing Surveys \textbf{46}(4), 44:1--44:37 (2014).
\newblock \doi{10.1145/2523813}

\bibitem{garg2022atc}
Garg, S., Balakrishnan, S., Lipton, Z.C., Neyshabur, B., Sedghi, H.: Leveraging
  unlabeled data to predict out-of-distribution performance.
\newblock In: International Conference on Learning Representations (ICLR)
  (2022).
\newblock \urlprefix\url{https://openreview.net/forum?id=o_HsiMPYh_x}

\bibitem{gouveia2020towards}
Gouveia, A., Correia, M.: Towards quantum-enhanced machine learning for network
  intrusion detection.
\newblock In: 2020 IEEE 19th International Symposium on Network Computing and
  Applications (NCA) (2020).
\newblock \doi{10.1109/NCA51143.2020.9306691}

\bibitem{gretton2012kernel}
Gretton, A., Borgwardt, K.M., Rasch, M.J., Sch{\"o}lkopf, B., Smola, A.: A
  kernel two-sample test.
\newblock Journal of Machine Learning Research \textbf{13}, 723--773 (2012)

\bibitem{guillory2021doc}
Guillory, D., Shankar, V., Ebrahimi, S., Darrell, T., Schmidt, L.: Predicting
  with confidence on unseen distributions.
\newblock In: Proceedings of the IEEE/CVF International Conference on Computer
  Vision (ICCV), pp. 1134--1144 (2021).
\newblock \doi{10.1109/ICCV48922.2021.00117}

\bibitem{guissouma2023continuous}
Guissouma, H., Zink, M., Sax, E.: Continuous safety assessment of updated
  supervised learning models in shadow mode.
\newblock In: 2023 IEEE 20th International Conference on Software Architecture
  Companion (ICSA-C) (2023).
\newblock \doi{10.1109/ICSA-C57050.2023.00069}

\bibitem{han2024model}
Han, E., Huang, C., Wang, K.: Model assessment and selection under temporal
  distribution shift.
\newblock In: Proceedings of the 41st International Conference on Machine
  Learning (ICML), \emph{Proceedings of Machine Learning Research}, vol. 235,
  pp. 17374--17392 (2024)

\bibitem{havlicek2019supervised}
Havl{\'i}{\v{c}}ek, V., C{\'o}rcoles, A.D., Temme, K., Harrow, A.W., Kandala,
  A., Chow, J.M., Gambetta, J.M.: Supervised learning with quantum-enhanced
  feature spaces.
\newblock Nature \textbf{567}, 209--212 (2019).
\newblock \doi{10.1038/s41586-019-0980-2}

\bibitem{holm1979simple}
Holm, S.: A simple sequentially rejective multiple test procedure.
\newblock Scandinavian Journal of Statistics \textbf{6}(2), 65--70 (1979)

\bibitem{howard2021timeuniform}
Howard, S.R., Ramdas, A., McAuliffe, J., Sekhon, J.: Time-uniform,
  nonparametric, nonasymptotic confidence sequences.
\newblock The Annals of Statistics \textbf{49}(2), 1055--1080 (2021).
\newblock \doi{10.1214/20-AOS1991}

\bibitem{huang2021power}
Huang, H.Y., Broughton, M., Mohseni, M., Babbush, R., Boixo, S., Neven, H.,
  McClean, J.R.: Power of data in quantum machine learning.
\newblock Nature Communications \textbf{12}, 2631 (2021).
\newblock \doi{10.1038/s41467-021-22539-9}

\bibitem{jacobs2022emperor}
Jacobs, A.S., Beltiukov, R., Willinger, W., Ferreira, R.A., Gupta, A.,
  Granville, L.Z.: {AI/ML} for network security: The emperor has no clothes.
\newblock In: Proceedings of the 2022 ACM SIGSAC Conference on Computer and
  Communications Security (CCS '22), pp. 1537--1551. ACM (2022).
\newblock \doi{10.1145/3548606.3560609}

\bibitem{jiang2022disagreement}
Jiang, Y., Nagarajan, V., Baek, C., Kolter, J.Z.: Assessing generalization of
  {SGD} via disagreement.
\newblock In: International Conference on Learning Representations (ICLR)
  (2022).
\newblock \urlprefix\url{https://openreview.net/forum?id=WvOGCEAQhxl}

\bibitem{jordaney2017transcend}
Jordaney, R., Sharad, K., Dash, S.K., Wang, Z., Papini, D., Nouretdinov, I.,
  Cavallaro, L.: {Transcend}: Detecting concept drift in malware classification
  models.
\newblock In: 26th USENIX Security Symposium (USENIX Security 17), pp. 625--642
  (2017)

\bibitem{kaissar2026enhancing}
Kaissar, A., Nassif, A.B., Bouridane, A.: Enhancing network intrusion detection
  with quantum machine learning: A comprehensive survey of methods, metrics,
  and applications.
\newblock Future Internet \textbf{18}(5), 234 (2026).
\newblock \doi{10.3390/fi18050234}

\bibitem{kan2021labelless}
Kan, Z., Pendlebury, F., Pierazzi, F., Cavallaro, L.: Investigating labelless
  drift adaptation for malware detection.
\newblock In: Proceedings of the 14th ACM Workshop on Artificial Intelligence
  and Security (AISec '21), pp. 123--134. ACM (2021).
\newblock \doi{10.1145/3474369.3486873}

\bibitem{karimi2021online}
Karimi, M.R., G{\"u}rel, N.M., Karla{\v{s}}, B., Rausch, J., Zhang, C., Krause,
  A.: Online active model selection for pre-trained classifiers.
\newblock In: Proceedings of the 24th International Conference on Artificial
  Intelligence and Statistics (AISTATS), \emph{Proceedings of Machine Learning
  Research}, vol. 130, pp. 307--315 (2021)

\bibitem{kirkpatrick2017overcoming}
Kirkpatrick, J., Pascanu, R., Rabinowitz, N., Veness, J., Desjardins, G., Rusu,
  A.A., Milan, K., Quan, J., Ramalho, T., Grabska-Barwinska, A., Hassabis, D.,
  Clopath, C., Kumaran, D., Hadsell, R.: Overcoming catastrophic forgetting in
  neural networks.
\newblock Proceedings of the National Academy of Sciences \textbf{114}(13),
  3521--3526 (2017).
\newblock \doi{10.1073/pnas.1611835114}

\bibitem{komorniczak2022ensemble}
Komorniczak, J., Zyblewski, P., Ksieniewicz, P.: Statistical drift detection
  ensemble for batch processing of data streams.
\newblock Knowledge-Based Systems \textbf{252}, 109380 (2022).
\newblock \doi{10.1016/j.knosys.2022.109380}

\bibitem{kossen2021active}
Kossen, J., Farquhar, S., Gal, Y., Rainforth, T.: Active testing:
  Sample-efficient model evaluation.
\newblock In: Proceedings of the 38th International Conference on Machine
  Learning (ICML), \emph{Proceedings of Machine Learning Research}, vol. 139,
  pp. 5753--5763 (2021)

\bibitem{krawczyk2019adaptive}
Krawczyk, B., Cano, A.: Adaptive ensemble active learning for drifting data
  stream mining.
\newblock In: Proceedings of the Twenty-Eighth International Joint Conference
  on Artificial Intelligence (IJCAI), pp. 2763--2771 (2019).
\newblock \doi{10.24963/ijcai.2019/383}

\bibitem{kreuzberger2023mlops}
Kreuzberger, D., K{\"u}hl, N., Hirschl, S.: Machine learning operations
  ({MLOps}): Overview, definition, and architecture.
\newblock IEEE Access \textbf{11}, 31866--31879 (2023).
\newblock \doi{10.1109/ACCESS.2023.3262138}

\bibitem{lanvin2022errors}
Lanvin, M., Gimenez, P.F., Han, Y., Majorczyk, F., M{\'e}, L., Totel, {\'E}.:
  Errors in the {CICIDS2017} dataset and the significant differences in
  detection performances it makes.
\newblock In: Risks and Security of Internet and Systems: 17th International
  Conference, CRiSIS 2022, Revised Selected Papers, \emph{Lecture Notes in
  Computer Science}, vol. 13857, pp. 18--33. Springer (2023).
\newblock \doi{10.1007/978-3-031-31108-6_2}

\bibitem{lin1991divergence}
Lin, J.: Divergence measures based on the {Shannon} entropy.
\newblock IEEE Transactions on Information Theory \textbf{37}(1), 145--151
  (1991).
\newblock \doi{10.1109/18.61115}

\bibitem{liu2021comprehensive}
Liu, W., Zhang, H., Ding, Z., Liu, Q., Zhu, C.: A comprehensive active learning
  method for multiclass imbalanced data streams with concept drift.
\newblock Knowledge-Based Systems \textbf{215}, 106778 (2021).
\newblock \doi{10.1016/j.knosys.2021.106778}

\bibitem{lu2019learning}
Lu, J., Liu, A., Dong, F., Gu, F., Gama, J., Zhang, G.: Learning under concept
  drift: A review.
\newblock IEEE Transactions on Knowledge and Data Engineering \textbf{31}(12),
  2346--2363 (2019).
\newblock \doi{10.1109/TKDE.2018.2876857}

\bibitem{lukats2024benchmark}
Lukats, D., Zielinski, O., Hahn, A., Stahl, F.: A benchmark and survey of fully
  unsupervised concept drift detectors on real-world data streams.
\newblock International Journal of Data Science and Analytics  (2024).
\newblock \doi{10.1007/s41060-024-00620-y}

\bibitem{mahadevan2024cost}
Mahadevan, A., Mathioudakis, M.: Cost-aware retraining for machine learning.
\newblock Knowledge-Based Systems \textbf{293}, 111610 (2024).
\newblock \doi{10.1016/j.knosys.2024.111610}

\bibitem{mccloskey1989catastrophic}
McCloskey, M., Cohen, N.J.: Catastrophic interference in connectionist
  networks: The sequential learning problem.
\newblock In: Psychology of Learning and Motivation, vol.~24, pp. 109--165.
  Academic Press (1989)

\bibitem{mcnemar1947note}
McNemar, Q.: Note on the sampling error of the difference between correlated
  proportions or percentages.
\newblock Psychometrika \textbf{12}(2), 153--157 (1947).
\newblock \doi{10.1007/BF02295996}

\bibitem{moustafa2015unsw}
Moustafa, N., Slay, J.: {UNSW-NB15}: a comprehensive data set for network
  intrusion detection systems ({UNSW-NB15} network data set).
\newblock In: 2015 Military Communications and Information Systems Conference
  (MilCIS) (2015).
\newblock \doi{10.1109/MilCIS.2015.7348942}

\bibitem{nath2007champion}
Nath, S.V.: Champion-challenger based predictive model selection.
\newblock In: Proceedings of IEEE SoutheastCon (2007).
\newblock \doi{10.1109/SECON.2007.342897}

\bibitem{nguyen2023tsids}
Nguyen, H., Kashef, R.: {TS-IDS}: Traffic-aware self-supervised learning for
  {IoT} network intrusion detection.
\newblock Knowledge-Based Systems \textbf{279}, 110966 (2023).
\newblock \doi{10.1016/j.knosys.2023.110966}

\bibitem{nicesio2023quantum}
Nicesio, O.K., Leal, A.G., Gava, V.L.: Quantum machine learning for network
  intrusion detection systems, a systematic literature review.
\newblock In: 2023 IEEE 2nd International Conference on AI in Cybersecurity
  (ICAIC) (2023).
\newblock \doi{10.1109/ICAIC57335.2023.10044125}

\bibitem{okanovic2025models}
Okanovic, P., Kirsch, A., Kasper, J., Hoefler, T., Krause, A., G{\"u}rel, N.M.:
  All models are wrong, some are useful: Model selection with limited labels.
\newblock In: Proceedings of the 28th International Conference on Artificial
  Intelligence and Statistics (AISTATS), \emph{Proceedings of Machine Learning
  Research}, vol. 258, pp. 2035--2043 (2025)

\bibitem{pederzoli2025noctowl}
Pederzoli, S., Paganelli, M., Contalbo, M.L., Benassi, R., Tiano, D., Iannucci,
  S., Guerra, F.: {NOCTOWL}: Adaptive tree-based model for network anomaly
  detection under delayed and sampled label availability.
\newblock IEEE Access \textbf{13}, 197899--197911 (2025).
\newblock \doi{10.1109/ACCESS.2025.3633419}

\bibitem{pendlebury2019tesseract}
Pendlebury, F., Pierazzi, F., Jordaney, R., Kinder, J., Cavallaro, L.:
  {TESSERACT}: Eliminating experimental bias in malware classification across
  space and time.
\newblock In: 28th USENIX Security Symposium (USENIX Security 19), pp. 729--746
  (2019)

\bibitem{raab2020reactive}
Raab, C., Heusinger, M., Schleif, F.M.: Reactive soft prototype computing for
  concept drift streams.
\newblock Neurocomputing \textbf{416}, 340--351 (2020).
\newblock \doi{10.1016/j.neucom.2019.11.111}

\bibitem{rabanser2019failing}
Rabanser, S., G{\"u}nnemann, S., Lipton, Z.C.: Failing loudly: An empirical
  study of methods for detecting dataset shift.
\newblock In: Advances in Neural Information Processing Systems (NeurIPS)
  (2019).
\newblock ArXiv:1810.11953

\bibitem{regol2025retrain}
Regol, F., Schwinn, L., Sprague, K., Coates, M., Markovich, T.: When to retrain
  a machine learning model.
\newblock In: Proceedings of the 42nd International Conference on Machine
  Learning (ICML), \emph{Proceedings of Machine Learning Research}, vol. 267,
  pp. 51369--51404 (2025)

\bibitem{rizzo2016energy}
Rizzo, M.L., Sz{\'e}kely, G.J.: Energy distance.
\newblock WIREs Computational Statistics \textbf{8}(1), 27--38 (2016).
\newblock \doi{10.1002/wics.1375}

\bibitem{rosenfeld2023dis2}
Rosenfeld, E., Garg, S.: {(Almost)} provable error bounds under distribution
  shift via disagreement discrepancy.
\newblock In: Advances in Neural Information Processing Systems (NeurIPS), pp.
  28761--28784 (2023).
\newblock
  \urlprefix\url{https://papers.nips.cc/paper_files/paper/2023/hash/5bacb12bf81e98e2ee0eed953a23c656-Abstract-Conference.html}

\bibitem{roshan2018adaptive}
Roshan, S., Miche, Y., Akusok, A., Lendasse, A.: Adaptive and online network
  intrusion detection system using clustering and extreme learning machines.
\newblock Journal of the Franklin Institute \textbf{355}(4), 1752--1779 (2018).
\newblock \doi{10.1016/j.jfranklin.2017.06.006}

\bibitem{sawade2012active}
Sawade, C., Landwehr, N., Scheffer, T.: Active comparison of prediction models.
\newblock In: Advances in Neural Information Processing Systems 25 (NeurIPS),
  pp. 1763--1771 (2012)

\bibitem{schnabel2025quantum}
Schnabel, J., Roth, M.: Quantum kernel methods under scrutiny: A benchmarking
  study.
\newblock Quantum Machine Intelligence \textbf{7}(1), 58 (2025).
\newblock \doi{10.1007/s42484-025-00273-5}

\bibitem{schuirmann1987comparison}
Schuirmann, D.J.: A comparison of the two one-sided tests procedure and the
  power approach for assessing the equivalence of average bioavailability.
\newblock Journal of Pharmacokinetics and Biopharmaceutics \textbf{15}(6),
  657--680 (1987).
\newblock \doi{10.1007/BF01068419}

\bibitem{schuld2019quantum}
Schuld, M., Killoran, N.: Quantum machine learning in feature hilbert spaces.
\newblock Physical Review Letters \textbf{122}, 040504 (2019).
\newblock \doi{10.1103/PhysRevLett.122.040504}

\bibitem{sculley2015hidden}
Sculley, D., Holt, G., Golovin, D., Davydov, E., Phillips, T., Ebner, D.,
  Chaudhary, V., Young, M., Crespo, J.F., Dennison, D.: Hidden technical debt
  in machine learning systems.
\newblock In: Advances in Neural Information Processing Systems (NeurIPS)
  (2015)

\bibitem{sharafaldin2018toward}
Sharafaldin, I., Lashkari, A.H., Ghorbani, A.A.: Toward generating a new
  intrusion detection dataset and intrusion traffic characterization.
\newblock In: Proceedings of the 4th International Conference on Information
  Systems Security and Privacy (ICISSP), pp. 108--116 (2018)

\bibitem{shyaa2024evolving}
Shyaa, M.A., Ibrahim, N.F., Zainol, Z., Abdullah, R., Anbar, M., Alzubaidi, L.:
  Evolving cybersecurity frontiers: A comprehensive survey on concept drift and
  feature dynamics aware machine and deep learning in intrusion detection
  systems.
\newblock Engineering Applications of Artificial Intelligence \textbf{137},
  109143 (2024).
\newblock \doi{10.1016/j.engappai.2024.109143}

\bibitem{shyaa2026igpcmsos}
Shyaa, M.A., Ibrahim, N.F., Zainol, Z.B., Abdullah, R., Anbar, M., Alzubaidi,
  L.: {IGPC-MSOS}: A knowledge-preserving transfer learning framework with
  dynamic mode-switching for handling concept drift in network intrusion
  detection systems.
\newblock Knowledge-Based Systems \textbf{337}, 115361 (2026).
\newblock \doi{10.1016/j.knosys.2026.115361}

\bibitem{sommer2010outside}
Sommer, R., Paxson, V.: Outside the closed world: On using machine learning for
  network intrusion detection.
\newblock In: 2010 IEEE Symposium on Security and Privacy (S\&P), pp. 305--316.
  IEEE (2010).
\newblock \doi{10.1109/SP.2010.25}

\bibitem{szekely2013energy}
Sz{\'e}kely, G.J., Rizzo, M.L.: Energy statistics: A class of statistics based
  on distances.
\newblock Journal of Statistical Planning and Inference \textbf{143}(8),
  1249--1272 (2013).
\newblock \doi{10.1016/j.jspi.2013.03.018}

\bibitem{thanasilp2024concentration}
Thanasilp, S., Wang, S., Cerezo, M., Holmes, Z.: Exponential concentration in
  quantum kernel methods.
\newblock Nature Communications \textbf{15}, 5200 (2024).
\newblock \doi{10.1038/s41467-024-49287-w}

\bibitem{waudbysmith2023betting}
Waudby-Smith, I., Ramdas, A.: Estimating means of bounded random variables by
  betting.
\newblock Journal of the Royal Statistical Society Series B: Statistical
  Methodology \textbf{86}(1), 1--27 (2024).
\newblock \doi{10.1093/jrsssb/qkad009}

\bibitem{yang2021cade}
Yang, L., Guo, W., Hao, Q., Ciptadi, A., Ahmadzadeh, A., Xing, X., Wang, G.:
  {CADE}: Detecting and explaining concept drift samples for security
  applications.
\newblock In: 30th USENIX Security Symposium (USENIX Security 21), pp.
  2327--2344 (2021)

\bibitem{yuan2026adawu}
Yuan, J., Yang, Y., Wang, M.: {ADAWU-IDS}: A dynamic adaptive weight update
  mechanism for concept drift-aware network intrusion detection systems.
\newblock Journal of King Saud University -- Computer and Information Sciences
  \textbf{38}, 546 (2026).
\newblock \doi{10.1007/s44443-026-00964-4}

\bibitem{zhang2024caravan}
Zhang, Q., Imran, A., Bardhi, E., Swamy, T., Zhang, N., Shahbaz, M., Olukotun,
  K.: {CARAVAN}: Practical online learning of in-network {ML} models with
  labeling agents.
\newblock In: 18th USENIX Symposium on Operating Systems Design and
  Implementation (OSDI 24), pp. 325--345. USENIX Association (2024).
\newblock
  \urlprefix\url{https://www.usenix.org/conference/osdi24/presentation/zhang-qizheng}

\bibitem{zhang2023survey}
Zhang, W., Deng, L., Zhang, L., Wu, D.: A survey on negative transfer.
\newblock IEEE/CAA Journal of Automatica Sinica \textbf{10}(2), 305--329
  (2023).
\newblock \doi{10.1109/JAS.2022.106004}

\bibitem{zhang2025ssf}
Zhang, X., Zhao, R., Jiang, Z., Chen, H., Ding, Y., Ngai, E.C.H., Yang, S.H.:
  Continual learning with strategic selection and forgetting for network
  intrusion detection.
\newblock In: IEEE INFOCOM 2025 -- IEEE Conference on Computer Communications,
  pp. 1--10. IEEE (2025).
\newblock \doi{10.1109/INFOCOM55648.2025.11044615}

\bibitem{zliobaite2015cost}
{\v{Z}}liobait{\.e}, I., Budka, M., Stahl, F.: Towards cost-sensitive
  adaptation: When is it worth updating your predictive model?
\newblock Neurocomputing \textbf{150}, 240--249 (2015).
\newblock \doi{10.1016/j.neucom.2014.05.084}

\bibitem{zoghi2024unsw}
Zoghi, Z., Serpen, G.: {UNSW-NB15} computer security dataset: Analysis through
  visualization.
\newblock Security and Privacy \textbf{7}(1), e331 (2024).
\newblock \doi{10.1002/spy2.331}

\end{thebibliography}
